\documentclass{elsart}

\usepackage{epsfig,subfigure,latexsym,amsmath}

\begin{document}

\def\Journal#1#2#3#4{{#1} {\bf #2}, #3 (#4)}
\def\RPP{{Rep. Prog. Phys.}}
\def\PRC{{Phys. Rev. C}}
\def\FP{{Foundations of Physics}}
\def\ZPA{{Z. Phys. A}}
\def\NPA{{Nucl. Phys. A}}
\def\JPG{{J. Phys. G. Nucl. Part}}
\def\PRL{{Phys.Rev.Letts.}}
\def\PRpt{{Phys. Report.}}
\def\PLB{{Phys. Letts. B}}
\def\AP{{Ann. Phys (N.Y.)}}
\def\EPJA{{Eur. Phys. J.A}}
\def\NP{{Nucl. Phys}}
\def\ZP{{Z. Phys}}
\def\RMP{{Rev. Mod. Phys}}

\input epsf

\begin{frontmatter}

\title{Mapping exchange in relativistic Hartree-Fock}

\author[indo]{A. Sulaksono}
\author[lanl]{T. B{\"u}rvenich}
\author[fra]{J.~A. Maruhn}
\author[er]{P.--G. Reinhard}
\author[fra]{W. Greiner}
\address[indo]{Jurusan fisika, FMIPA, Universitas Indonesia, Depok 16424, Indonesia}
\address[lanl]{Theoretical Division, Los Alamos National Laboratory, Los Alamos, New Mexico 87544}
\address[fra]{Institut f\"ur Theoretische Physik, Universit\"at
Frankfurt, Robert-Mayer-Str. 10, D-60324 Frankfurt, Germany}
\address[er]{Institut
f\"ur Theoretische Physik II, Universit\"at Erlangen-N\"urnberg, Staudtstrasse 7, D-91058 Erlangen, Germany}

\begin{abstract}
We show that formally for the standard ansatz relativistic
point-coupling mean-field (RMF-PC) model a lagrangian density
${\mathcal L}$ is not equivalent in Hartree and Hartree-Fock approximations. The
equivalency can be achieved only if we use a ``complete'' ansatz at the
cost of introducing new parameters in the model. An
approximate treatment of the exchange terms from standard RMF-PC
indicates that these effects cannot be easily, if at all, absorbed by a
Dirac-Hartree approximation.
\end{abstract}

\end{frontmatter}        

\section{Introduction}
Relativistic mean-field  models, which describe the nucleus as a 
system of Dirac nucleons that interact with each other by exchanging mean
meson fields, have been successful in describing nuclear matter and ground 
state properties of finite
nuclei~\cite{pgr,rufa,pg,sharma,Gambhir,Boersma,Lalazissis,Klemens}. 
Applications include
the calculation of deformed
nuclei~\cite{Klemens,bur,cornel,Kudling}, odd
nuclei~\cite{Klemens}, nuclei at the
driplines~\cite{Naza}, the potential
energy surfaces of heavy nuclei~\cite{Klemens}, and the prediction of
superheavies ~\cite{Rutz2,Klemens,Bender3}.  Another variant is the point-coupling model~\cite{sur,burven,niko}. 
The difference between this model and
the Walecka model is the replacement of the mesonic potentials of
the Walecka model by explicitly density-dependent
potentials. 
Nikolaus, Hoch, and Madland~\cite{niko} used the
Hartree form of this model to calculate some observables of finite
nuclei and nuclear matter. They obtained similar
predictive power as the Walecka model, but used different weights and
observables from Ref.~\cite{pgr} to obtain their parameter set.  In
Ref.~\cite{sur}, Rusnak and Furnstahl have shown the
profitability of applying the concepts of effective field theory to the
point-coupling model. A more recent discussion about the point-coupling model can be found in 
Ref.~\cite{burven,buervi}. It consists of a careful way to obtain a new parameter
set for the point-coupling model which is likely to correspond to
the optimal minimum by combining different
methods for the $\chi^2$ minimization. This parameter set was applied to a wide area in nuclear
theory, from spherical finite nuclei, axially deformed nuclei, superheavy
elements, potential energy surfaces, nuclear matter, neutron matter,
up to exotic nuclei. It could be shown that point-coupling models can
deliver a predictive power for finite nuclei comparable to models
employing meson exchange. On the other hand, especially concerning
isovector properties, the point-coupling model suffers from the same
unresolved systematic deviations.

From the success of the RMF (Walecka and point-coupling) models, we suppose that the
exchange effect is already absorbed effectively in the coupling
constants of the model due to the fitting procedure, but some
calculations which take into account the exchange explicitly  by using
the linear Walecka model have shown that this does not seem to be the
case~\cite{nley,schmd}. Understanding this situation seems
necessary and interesting. Unfortunately, due to the finite range of the mesons
and the mesonic nonlinearity of the RMF model, it is difficult to
expand this model in the above direction. This problem does not
appear if we
use the point-coupling model~\cite{niko}, because in this model the
linear and nonlinear terms are explicitly density dependent. Therefore
in this paper, we choose the
point-coupling model for a first study of the role of the exchange
effects. 

We would like to stress that we regard this work as a model study.
The aim is to check the principle effects of exchange terms
as compared to an effective Hartree theory.
This is done for the two-body point-coupling terms of the
standard RMF model. Higher order couplings are treated without exchange
and the pseudoscalar (pionic) channel is omitted for the time being.

In the following, we will use the phrase {\em strict Hartree} for an ansatz that
is taken from the start in the Hartree approximation. We will use {\em
Hartree-Fock} for a model in which the exchange terms 
are treated explicitly. If this model can be transformed into a form consisting
of local currents and densities only (no matter how complicated they are), we
will call the result of this rewriting an {\em effective Hartree} theory.
Mapping formally this general form onto a Dirac-like structure leads to the {\em
Dirac-Hartree} theory.

\section{The formulation of Hartree-Fock Point-Coupling Model}

\subsection{Basic one-particle operators and densities}

We beginn by defining the basic Dirac \& isovector matrices $\Gamma_j$.
There are two sets
$$
  \Gamma_j\in
  \left\{
   \begin{array}{lcl}
     {\rm H} &=& \{1,\gamma_\mu,\gamma_\mu\vec{\tau}\} \\[4pt]
     {\rm HF} &=&
  H\cup\{\vec{\tau},\sigma_{\mu\nu},\sigma_{\mu\nu}\vec{\tau},
               \gamma_5,\gamma_5\vec{\tau},\gamma_5\gamma_\mu,
               \gamma_5\gamma_\mu\vec{\tau}\} 
   \end{array}
   \right .
$$
The ``Hartree'' set H covers the currents initially given in the
RMF-PC Lagrangian. The ``Hartree-Fock'' set HF covers all conceivable
Dirac matrices as they are produced by the Fierz transformation
\cite{GM}. The currents are defined as:
$$
  \hat{\mathcal J}_j
  =
  \hat{\bar{\psi}}\Gamma_j\hat{\psi}
  \quad,\quad
  \hat{\mathcal A}_{j\mu}
  =
  \partial_\mu\hat{\mathcal J}_j
  \quad,\quad
  \hat{\mathcal B}_{j\mu}
  =
  \hat{\bar{\psi}}
  \frac{\stackrel{\leftrightarrow}{\partial_\mu}}{2i}
  \Gamma_j\hat{\psi}
  \quad,\quad
  \hat{\mathcal C}_j
  =
  \partial_\mu\hat{\bar{\psi}}\Gamma_j\partial^\mu\hat{\psi}
  \quad,
$$
where 
$\stackrel{\leftrightarrow}{\partial_\mu} = 
 \stackrel{\leftarrow}{\partial_\mu}-\stackrel{\rightarrow}{\partial_\mu}$.
These are the current operators from which actually only 
$\hat{\mathcal J}_j$ is used at operator level. The associated currents are 
the expectation values
\begin{eqnarray}
  {\mathcal J}_j
  &=&
  \sum_\alpha \bar{\Psi}_{\alpha}\Gamma_{j}\Psi_{\alpha}  \quad, \quad
 \quad
  {\mathcal A}_{j\mu}
  =
  \partial_\mu{\mathcal J}_j
  \quad,\quad\nonumber\\
  {\mathcal B}_{j\mu}
  &=&
  \sum_\alpha \bar{\Psi}_{\alpha}
  \frac{\stackrel{\leftrightarrow}{\partial_\mu}}{2i}
  \Gamma_j \Psi_{\alpha}
  \quad,\quad
  {\mathcal C}_j
  =
   \sum_\alpha \partial_\mu\bar{\Psi}_{\alpha}\Gamma_j\partial^\mu\Psi_{\alpha}
  \quad, \label{pis1}
\end{eqnarray}
where the sum is running over occupied particle states only 
(anti-particles are neglected = no-sea approximation). 

\subsection{The Lagrangian}
Starting point is the Lagrangian
\begin{subequations}
\label{eq:lagr}
\begin{eqnarray}
\hat{\mathcal L}
  &=&
 \hat{\mathcal L}_{\rm free} + \hat{\mathcal L}_{\rm L}^{PC} + \hat{\mathcal L}_{\rm
NL}^{PC} + \hat{\mathcal L}_{\rm coul},
\\
  \hat{\mathcal L}_{\rm L}^{ PC}
  &=&
  \frac{1}{2}\sum_{j\in{\rm H}}[
    \alpha_j\hat{\mathcal J}_j^2
    +
    \delta_j\hat{\mathcal A}_j^2  ],
\\
   \hat{\mathcal L}_{\rm NL}^{PC}
  &=& 
  \frac{1}{3}\beta_S\hat{\mathcal J}_S^3+\frac{1}{4}\gamma_S\hat{\mathcal J}_S^4.
\end{eqnarray}
\end{subequations}
At this point, replacing the current operators by their expectation
values leads to a strict Hartree theory, as is mostly applied 
to nuclear structure problems.

But we are going here for a Hartree-Fock treatment. 
To this end we chose a Slater determinant $| 0\rangle$ as the ground state
with $A$ occupied single-nucleon levels (in the positive energy
sector). It is defined by
\begin{eqnarray}
 b_{i}^{\dagger} | 0\rangle = 0  \quad \forall  \quad i \leq F,
\end{eqnarray}
($F$ indicates the fermi surface).
We compute the expectation value of the Lagrangian (\ref{eq:lagr}) 
in standard manner and employ Fierz transformations to express
exchange terms through simple currents.This yields
\begin{subequations}
\label{pos13}
\begin{eqnarray} 
  \langle 0|{\mathcal{L}}_{\rm L}^{PC}|0\rangle 
  &=& 
  {\mathcal L}_{\rm L}^{ PC} 
  -
  {\mathcal L}_{\rm Lx}^{ PC} 
  \quad,
\\
  {\mathcal L}_{\rm L}^{ PC} 
  &=& 
  \sum_{i\in{\rm H}}
  \Big[\frac{1}{2} ~\alpha_i~ {\mathcal J}_{i }^{2}+\frac{1}{2}~
  \delta_i~ {\mathcal A}_{i }^{2}\Big],
\\
  {\mathcal L}_{\rm Lx}^{ PC} 
  &=& 
   \sum_{i\in{\rm H}} \sum_{j\in{\rm HF}} \Big[ C_{ji}~ \Big(\frac{1}{2}~\alpha_i~ {\mathcal J}_{j}^{2}
  +
  \delta_i ~( \frac{1}{4} {\mathcal A}_{j }^{2}- {\mathcal B}_{j}^{2}+ {\mathcal J}_{j}~
  {\mathcal C}^{j})\Big)\Big].
\end{eqnarray}
\end{subequations}
The constants $ C_{ji}$ stem from the Fierz transformation, their
explicit form can be found in table~\ref{tab1}. 

Thus we have performed - by virtue of the Fierz transformation - a
mapping from a Hartree Fock model to an effective Hartree model for
all linear terms.  The expectation value of non-linear term
$\hat{\mathcal{L}}_{\rm NL}^{PC}$ has been calculated in
Refs.~\cite{ma,burven}. The results are quite lengthy, therefore it is
not useful to consider them explicitly here. The exchange terms
emerging from the nonlinear part of the Lagrangian form a large
variety of density-mixing terms, involving also products of isoscalar
and isovector densities. This leads to modifications in both the
isoscalar and the isovector channel of the effective interaction
without introducing new parameters. Another feature of this result is
that the contributions from the nonlinear exchange terms cannot be
absorbed fully by a redefinition of the coupling constants into terms
of ${\mathcal L} _{\rm N L}^{PC }$ and they also cannot be considered
as small corrections.
\begin{table}
\centering
\begin{tabular}{|l|c|c|c|}
\hline $C_{ji}$ & $\Gamma_{i}$=1 &$\Gamma_{i}$ =$\gamma_{\mu}$ & $\Gamma_{i}$=$\gamma_{\mu} \vec{\tau}$ \\\hline
$\Gamma_{j}$ =1 & $\frac{1}{8}$  & $\frac{4}{8}$  & $\frac{12}{8}$ \\\hline
$\Gamma_{j}$=$\vec{\tau}$ &  $\frac{1}{8}$ & $\frac{4}{8}$  & -$\frac{4}{8}$\\\hline
$\Gamma_{j}$=$\gamma_{\mu}$ & $\frac{1}{8}$ & -$\frac{2}{8}$ & -$\frac{6}{8}$   \\\hline
$\Gamma_{j}$=$\gamma_{\mu} \vec{\tau}$ & $\frac{1}{8}$  & -$\frac{2}{8}$  &
$\frac{2}{8}$  \\\hline
$\Gamma_{j}$=$\sigma_{\mu \nu} $ & $\frac{1}{16}$  & 0  &
0  \\\hline
$\Gamma_{j}$=$\sigma_{\mu \nu} \vec{\tau}$ & $\frac{1}{16}$  & 0  &
0  \\\hline
$\Gamma_{j}$=$\gamma_{5}$ & $\frac{1}{8}$ & -$\frac{4}{8}$ & -$\frac{12}{8}$
\\\hline
$\Gamma_{j}$=$\gamma_{5}\vec{\tau}$ & $\frac{1}{8}$ & -$\frac{4}{8}$ & $\frac{4}{8}$
\\\hline
$\Gamma_{j}$=$\gamma_{5}\gamma_{\mu} $ & $\frac{1}{8}$ & $\frac{2}{8}$ & $\frac{6}{8}$
\\\hline
$\Gamma_{j}$=$\gamma_{5}\gamma_{\mu}\vec{\tau} $ & $\frac{1}{8}$ & $\frac{2}{8}$ & -$\frac{2}{8}$
\\\hline
\end{tabular}\\
\caption {The $C_{ji}$ constants of equation (\ref{pos13}).\label{tab1}}
\end{table}

\subsection{The Hartree-Fock equations}
Variation of the effective Lagrangian density (\ref{pos13}) with
respect to the single-nucleon wavefunctions yields the effective
mean-field equations 
\begin{subequations}
\label{pos16}
\begin{eqnarray}
  0
  &=&
  \left[
   -i\gamma_\mu\partial^\mu+m_B
   + V^j\Gamma_j + U^{j\mu}\Gamma_j\partial_\mu
   + W^j\Gamma_j\partial^\mu\partial_\mu
  \right]\Psi_\alpha
\label{eq:dirac}
\\
  V^j\Gamma_j
  &=&
  \tilde{S}+\gamma_\alpha \tilde{V}^{\alpha}
\\
  U^{j\mu}\Gamma_j
  &=&
  U_{\mu}+\gamma^{\alpha}U_{\alpha \mu}
\\
  W^j\Gamma_j
  &=& 
  W^1+\gamma_{\alpha}W^{2 \alpha}
\\
  \tilde{S} 
  &=& 
  \alpha_a  {\mathcal J}_a - (\delta_a +\tilde{\delta}_a)
  \partial^{\mu} {\mathcal A}_{\mu a}+\frac{2}{3}\tilde{\delta}_a (i
  \partial^{\mu}  {\mathcal B}_{\mu a}+  {\mathcal C}_a)
\\
  \tilde{V}^{\alpha}
  &=&
  \alpha_b  {\mathcal J}_b^{\alpha} - (\delta_b +\tilde{\delta}_b)
  \partial^{\mu} {\mathcal A}_{\mu b}^{\alpha}+\frac{2}{3}\tilde{\delta}_b
  (i \partial^{\mu}  {\mathcal B}_{\mu b}^{\alpha}+  {\mathcal C}_b^{\alpha}),
\\
  U_{\mu} 
  &=& 
  \frac{4}{3} i \tilde{\delta}_a {\mathcal B}_{\mu a},~ ~ ~ ~ ~
  U_{\nu\mu}= \frac{4}{3} i \tilde{\delta}_b  {\mathcal B}_{\mu \nu b}, 
\\
  W_1
  &=& 
  -\frac{2}{3} \tilde{\delta}_a  {\mathcal J}_a, ~ ~ ~ ~ ~
  W_{1 \alpha}= -\frac{2}{3}\tilde{\delta}_b  {\mathcal J}_{b \alpha}.
\label{posi17}
\end{eqnarray}
\end{subequations}
A double index $a/b$ implies summations over
S,D/V,R. The constants in front of the densities in
Eq. (\ref{posi17}) (${\alpha}_{a/b}$, 
${\delta}_{a/b}$, $\tilde{\delta}_{a/b}$) are  functions of
the meson coupling constants and masses. The ${\delta}_{a/b}$,
$\tilde{\delta}_{a/b}$ constants reflect the combinations of Fierz constants $C_{ij}$, the meson coupling
constants and masses of the mesons from the derivative exchange terms of Eq. (\ref{pos13}) in the equation of motion. Because the formulas for
the constants are lengthy and do not give particular
insight, we do not show them.
$\tilde{S},\tilde{V}^{\alpha}$, $W^1,
 W^{2 \alpha}$ are of order $<$ 1 and $U_{\mu}, U_{\alpha \mu}$ are of
order $v$. We can see here that the $ {\mathcal C}_{i  }$ densities
give rise to the term with prefactor $W^j$ and the ${\mathcal B}_{i
\mu}$ the term with prefactor $U^{j \mu}$  in the equation of motion.

Equation (\ref{eq:dirac}) starts out like a Dirac equation and
continues with very unconventional terms $\propto\Gamma_j\partial_\mu$
and $\propto\Gamma_j\partial^\mu\partial_\mu$. The latter are
generated from the Fierz transformed exchange terms $\propto{\mathcal B}$
and ${\mathcal C}$ in Eq. (\ref{pos13}). These terms are awkward to
handle.  We would like to transform Eq. (\ref{eq:dirac}) into a more
standard form. This will be done in section \ref{sec:approx} by means
of a Gordon decomposition. Before that we first check the importance
of the new terms. This will be done in the following section.

From the discussion above it is clear that, in contrast to the claim
of Ref.~\cite{niko}, qualitatively, if a standard ansatz is used, a Lagrangian densitiy ${\mathcal L}^{HF}$ can
be determined in a relativistic Hartree-Fock sense that is not equivalent
to that determined in a relativistic Hartree sense, ${\mathcal
L}^{H}$. Of course, the Hartree or Hartree-Fock calculations 
belong to effective theories, where many details or ignored
corrections can be hidden in the form of the effective interaction and
in its fitted parameters. In this
sense, the models can only be compared after refitting the
parameters. Therefore, only quantitative calculations can actually
distinguish the predictive capabilities of the models. Before
embarking into the actual calculation, though, we present a rough
approximation that gives physical insight into the exchange contributions. 

\section{ Estimating the  Densities ${\mathcal B}_{j\mu}$ and ${\mathcal C}_{j}$}
\label{sec3}
For a first overview, we compare the order of magnitude of the energy
density contributions of each term in Eq. (\ref{pos13}) in the plane
wave (nuclear matter) limit.  The dominant contributions in the energy
density come from the terms with $\Gamma_j$ = 1, $\gamma_0$.  The
terms with $\Gamma_j$ = $\gamma_0$ show a clear order counting and are
easy to calculate. It is illustrative to examine their order of
magnitude. They will be represented by :
\begin{eqnarray*}
  \epsilon_1  
  &=& 
  -\frac{1}{2 }\alpha_V  {\mathcal J}_{V}^2
\\
  \epsilon_2  
  &=&
  \frac{1}{2 } \delta_V {\mathcal A}_{V}^2
\\
  \epsilon_{ex 1}  
  &=& 
  -\frac{c_v }{2} \alpha_V  {\mathcal J}_{V}^2
\\
  \epsilon_{ex 2}^1 
  &=& 
  c_v  \delta_V {\mathcal B}_{V}^{2}
\\
  &\approx& 
  -\frac{c_v }{4} \delta_V \sum_{\alpha \beta}[\bar{\Psi}_{\alpha}\gamma_0
  (\vec{\nabla}\Psi_{\alpha})-(\vec{\nabla}\bar{\Psi}_{\alpha})\gamma_0
  \Psi_{\alpha}].[\bar{\Psi}_{\beta}\gamma_0
  (\vec{\nabla}\Psi_{\beta})-(\vec{\nabla}\bar{\Psi}_{\beta})\gamma_0
  \Psi_{\beta}]
\\
  \epsilon_{ex 2}^2 
  &=&  
  c_v \delta_V {\mathcal J}_{V}~ {\mathcal C}_{V} \approx  c_v \delta_V 
  \sum_{\alpha \beta}(\bar{\Psi}_{\alpha}\gamma_0
  \Psi_{\alpha})(\vec{\nabla}\bar{\Psi}_{\beta}).\gamma_0
  (\vec{\nabla}\Psi_{\beta})
\end{eqnarray*}
$c_v$ is a constant due to the Fierz transformation. It is a function
of the coupling constants and fulfills $c_v$ $\leq$
1. Because the formula for $c_v$ is lengthy and does not give
particular insight, we do not show it here. In the last two equations
($\epsilon_{ex 2}^1$ and $\epsilon_{ex 2}^2$), we neglect the energy
dependent parts (retardation parts) by assuming that this contribution
is small. Then in the plane-wave approximation the order counting
which is shown in table \ref{pos15} is obtained.
\begin{table}
\centering
\begin{tabular}{|l|c|c|c|}
\hline Terms & expression & description & order\\\hline

\hline $\epsilon_1$  &  $\sim$ $-\frac{1}{2}\alpha_V \bar{\rho_0^2}$
& direct nonderivative term & $v^0$\\\hline

\hline $\epsilon_2$ &  $\sim$ 0 & direct derivative term & 0\\\hline

\hline $\epsilon_{ex 1}$ & $\sim$ $-\frac{1 }{2
}c_v \alpha_V \bar{\rho_0^2}$ & exchange nonderivative term & $v^0$ \\\hline

\hline $\epsilon_{ex 2}^1$ & $\sim$ $- c_v \delta_V \frac{3.4}{6.02}\bar{\rho_0^2}p_F^2 $ &exchange derivative term
 & $v^2$\\\hline

\hline $\epsilon_{ex 2}^2$ & $\sim$ $ c_v \delta_V  \frac{3.6}{6.02}\bar{\rho_0^2}p_F^2$ &exchange derivative term
 & $v^2$\\\hline
\end{tabular}\\
\caption {The order counting \label{pos15}}
\end{table}
We use 
$$
  \bar{\rho_0}
  \equiv 
  \frac{(2S+1)(2I+1)}{{(2 \pi)}^3}\int d^3k \bar{\Psi}_{k}\gamma_0 \Psi_{k}
$$ 
where S and I denotes spin and isospin, respectively, and
$p_F$ is the fermi momentum. If we define
$p_F/m_v = v$, every exchange term in the
derivative part gives a contribution of order $v^2$, where  $v <$ 1. Then in the nuclear matter limit, $\bar{\rho_i} \neq$ 0,
$\vec{\nabla}{\bar{\rho_i}}$ =0, $\mid \vec{\bar{ {\mathcal B}_i}}\mid/(m_B
\bar{\rho_i})$ is of order $v$ and $\mid \bar{ {\mathcal C}_i}\mid/(m_B^2
\bar{\rho_i})$ is of order $v^2$, while in finite nuclei, we know that 
$\rho_i \neq$ 0 and
$\vec{\nabla}{\rho_i}\neq $0, where $\mid
\vec{\nabla}\rho_i\mid/(m_i \rho_i)$ can be estimated  to be of 
order $v$ or maybe even smaller~\cite{niko} by comparing
the direct and derivative terms in the Hartree level of the
point-coupling model. Here the index 
$i$ denote V or S. If we consider the densities in finite nuclei as consisting of 
the nuclear matter value plus fluctuation corrections due to quantum effects, it seems reasonable to assume in finite nuclei that on the  average
$\mid \vec{ {\mathcal B}_i}\mid/(m_B \rho_i)$ is also of order $v$ and $\mid
 {\mathcal C}_i\mid/(m_B^2 \rho_i)$ of order $v^2$.  

In conclusion, it is reasonable to assume that the contribution from 
the derivative exchange terms could be similar or even a bit larger than the
direct derivative terms, so that they cannot be neglected if the
direct derivative terms, which are necessary for a reasonable
description of the nuclear surface, are kept.

It should be noted that 
even the nonderivative exchange terms are larger than the
derivative terms in order of magnitude, but their largest contributions (scalar-isoscalar, vector-isoscalar and
vector-isovector) can be absorbed into the nonderivative terms through
a redefinition of the coupling constants and refitting of the experimental
data. The crucial difference between Hartree and Hartree-Fock calculations of
the nonderivative terms is represented by the
scalar-isovector term, which is usually omitted in most Hartree models. In the Hartree-Fock approach
presented here, it emerges naturally without introducing an additional parameter. 

As a further check of our procedure, we show a transformed form of the $ {\mathcal B}^{
\sigma}$ terms through
the Gordon decomposition~\cite{greiner} by using the exact equation 
of motion in
Eq. (\ref{pos16}). Here we only calculate the scalar case; the
vector case is similar but not as simple. In the limit of small $v^2$ (for more details see Appendix (2)),
we have
\begin{eqnarray}
 {\mathcal B}^{ \sigma} \approx \sum_{\alpha} \Big[- \frac{1}{2}\partial_{\mu}(\bar{\Psi}_{\alpha}
\sigma^{\mu\sigma}\Psi_{\alpha})-\bar{m}^*\bar{\Psi}_{\alpha}\gamma^{\sigma}\Psi_{\alpha}
-\bar{V}^{\sigma}\bar{\Psi}_{\alpha} \Psi_{\alpha}\Big].\label{poses17}
\end{eqnarray}
The same result as above is obtained by using the equation of
motion in the form of a Dirac equation with scalar potential ($\bar{m}^*$-m)
and vector potential $\bar{V}^{\sigma}$. It
means that in the limit of small $v$ the $ {\mathcal B}^{ \sigma}$ can be
determined from the Dirac equation (second-order differential equation).

A similar situation holds for all  $ {\mathcal C}_{i  }$ and
$ {\mathcal B}_{i  \mu}$, so that we can say that the transformed
form of  the exact $ {\mathcal C}_{i }$ and
$ {\mathcal B}_{i \mu}$ consist of parts responsible for generating
effects beyond the Dirac equation. The order of magnitude
analysis above shows that these parts are small. Neglecting them
everywhere (via projecting onto a Dirac-Hartree 
structure) will lead to the Dirac equation but neglecting these
small parts means also neglecting the retardation and nonlocal
effects. This is what we will do in the next section, where we map the exact
Lagrangian including exchange onto a Dirac-Hartree structure.

\section{The Approximate Lagrange Density}
\label{sec:approx}
The aim of this chapter is to derive an effective Lagrangian which
leads to a Dirac equation without extra derivative terms, i.e., we will
formulate, by employing some approximations, a Dirac-Hartree model.

The approximate forms of 
the densities $ {\mathcal C}_{i  }$ and $ {\mathcal B}_{i  \mu}$
are not yet known. We expect to obtain from
them an approximate Lagrangian density (with the approximate forms of 
the densities $ {\mathcal C}_{i  }$ and $ {\mathcal B}_{i  \mu}$),
which should lead to a Dirac equation of the form
\begin{eqnarray}
[ \gamma_{\mu} \partial^{\mu}+i (m^*+\gamma_{\alpha}
V^{\alpha}+\sigma_{\alpha \beta} T^{\alpha \beta})]\Psi_{\alpha}=0,\label{pos20}
\end{eqnarray}  
where $m^*$, $V^{\alpha}$ and $T^{\alpha \beta}$ are real
functions. It implies a connection between the two types of functions:
$m^*$, $V^{\alpha}$, $T^{\alpha \beta}$ and $ {\mathcal J}_{i }$, $ {\mathcal
A}_{i\mu}$, $ {\mathcal C}^{j}$, $ {\mathcal B}_{j \mu}$. Therefore we will
try to obtain the $ {\mathcal C}^{j}$ and $ {\mathcal B}_{j \mu}$ as functions
of the $m^*$, $V^{\alpha}$ , and $T^{\alpha \beta}$.
This will be done in an iterative procedure starting from $m^*-m$, $V$, $T=0$
and generating succesively terms with increasing orders.
Finally, we use the approximate form of $m^*$, $V^{\alpha}$ , and
$T^{\alpha \beta}$ to calculate $ {\mathcal C}^{j}$ and $ {\mathcal B}_{j
\mu}$ and from these approximate densities $ {\mathcal C}^{j}$ and $ {\mathcal
B}_{j \mu}$ we obtain the approximate Lagrange density. These steps
will now be done explicitly for the scalar case ($\Gamma$ =1).

Starting point is the Gordon decomposition~\cite{greiner} of
Eq. (\ref{pos20}) which leads to
\begin{eqnarray}
\bar{\Psi}_{\alpha}[ \gamma_{\mu} \overleftarrow{\partial^{\mu}}-i (m^*+\gamma_{\alpha}
V^{\alpha}+\sigma_{\alpha \beta} T^{\alpha \beta})]\not a
\Psi_{\alpha}\nonumber\\-\bar{\Psi}_{\alpha}\not a [ \gamma_{\mu} \partial^{\mu}+i (m^*+\gamma_{\alpha}
V^{\alpha}+\sigma_{\alpha \beta} T^{\alpha \beta})]\Psi_{\alpha}=0,\label{pos21}
\end{eqnarray}
$\not{a} = a^{\mu} \gamma_{\mu}$ and $a^{\mu}$ is an abritrary
vector. Chosing $\not{a} = \gamma_{\mu}$ yields
\begin{eqnarray}
  {\mathcal B}_S
  &=&
  \frac{-i}{2}\Big(
    \bar{\Psi}_{\alpha} (\partial^{\nu}\Psi_{\alpha})-
    (\partial^{\nu}\bar{\Psi}_{\alpha})\Psi_{\alpha} \Big)
\nonumber\\
  &=& 
  -i[\partial_{\mu}(\bar{\Psi}_{\alpha} \sigma^{\mu\nu}\Psi_{\alpha})+2
  m^*\bar{\Psi}_{\alpha}\gamma^{\nu}\Psi_{\alpha}
  + 
  2V^{\nu}\bar{\Psi}_{\alpha} \Psi_{\alpha}+ 2 \epsilon^{\alpha \beta \nu
  \theta} T_{\alpha \beta} \bar{\Psi}_{\alpha} 
  \gamma_{\theta}\gamma_{5}\Psi_{\alpha} ],
\label{pos22}
\end{eqnarray}
From Eq. (\ref{pos20}) we deduce directly
\begin{eqnarray}
  {\mathcal C}_S
  &=&
  (\partial_{\mu}\bar{\Psi}_{\alpha})(\partial^{\mu}\Psi_{\alpha}) 
\nonumber\\
  &=&
  [m^{* 2 }\bar{\Psi}_{\alpha} \Psi_{\alpha}+ 2 m^{*}
  V^{\nu}\bar{\Psi}_{\alpha}\gamma_{\nu} \Psi_{\alpha}
  + 1/2\partial^{\mu}\partial_{\mu} (\bar{\Psi}_{\alpha}  \Psi_{\alpha})
\nonumber\\ 
  &&
  +V^{\nu}V_{\nu}\bar{\Psi}_{\alpha}
  \Psi_{\alpha}-\partial_{\nu}(V_{\mu} 
  \bar{\Psi}_{\alpha} \sigma^{\mu\nu}\Psi_{\alpha})+ \bar{\Psi}_{\alpha}
  f( T_{\alpha \beta}) \Psi_{\alpha}].
\label{pos23}
\end{eqnarray}
Here $f( T_{\alpha \beta})$ denotes a complicated function of
$T_{\alpha \beta}$ which is zero for $T_{\alpha \beta}$=0 and the
order of magnitude of this functions can be estimated as small
therefore it is not necessary to show this function explicitly.  These
are the desired expressions for ${\mathcal B}_S$ and ${\mathcal C}_S$ in terms
of $m^*$, $V$, and $T$. We would like to express them in terms of standard
currents. The construction for that proceeds as follows:
  
The iteration procedure leads to  ${m^{*}}= m_B+\tilde{\alpha}_S\rho_{S}(x)$+...,
   ${V^{\nu}}=\tilde{\alpha}_V J_{V}^{\nu}(x)$ +...,  and
  ${T^{\alpha \beta}}= ...$, where 
  $\tilde{\alpha}_S$ and
  $\tilde{\alpha}_V$  are effective
  coupling constants. They are functions of  $\alpha_S $,  $\alpha_V $ ,
  $\delta_S $ and $\delta_V$, which  come from the contribution of
  the direct and the exchange part in front of the nonderivative and 
  the derivative of the scalar and vector
  densities, respectively. In
  the above equations, ... denotes contributions from the derivative, 
  isovector, nonlinear
  and tensor correction parts. They will generate nonlinear terms in the Lagrangian density that consist of all
  possible combinations of the above mentioned densities and of all 
  possible order.  
  Now we obtain the
  approximate forms by neglecting ... (neglecting higher order terms)\\
 yielding
  \begin{subequations}
  \begin{eqnarray}
   {\mathcal B}_S
   &=&
    -i[\partial_{\mu}(\bar{\Psi}_{\alpha} \sigma^{\mu\nu}\Psi_{\alpha})+2
  (m_B+\tilde{\alpha}_S\rho_{S})\bar{\Psi}_{\alpha}\gamma^{\nu}\Psi_{\alpha}
  + 
  2\tilde{\alpha}_V J_{V}^{\nu} \bar{\Psi}_{\alpha} \Psi_{\alpha}],
  \\
   {\mathcal C}_S
   &=&
    [{(m_B+\tilde{\alpha}_S\rho_{S})}^{ 2 }\bar{\Psi}_{\alpha} \Psi_{\alpha}+ 2 (m_B+\tilde{\alpha}_S\rho_{S})\tilde{\alpha}_V J_{V}^{\nu}
  \bar{\Psi}_{\alpha}\gamma_{\nu} \Psi_{\alpha}
  + 1/2\partial^{\mu}\partial_{\mu} (\bar{\Psi}_{\alpha}  \Psi_{\alpha})
\nonumber\\ 
  &&
  +{(\tilde{\alpha}_V J_{V}^{\nu})}^2 \bar{\Psi}_{\alpha}
  \Psi_{\alpha}-\partial_{\nu}(\tilde{\alpha}_V J_{V \mu} 
  \bar{\Psi}_{\alpha} \sigma^{\mu\nu}\Psi_{\alpha})].
  \end{eqnarray}
  \end{subequations}

  Putting the pieces together yields
  \begin{eqnarray}
    \tilde{\mathcal L}_{S,Lx}^{PC(0)} 
    &=&  
    -\frac{1}{2}\partial_{\mu}\rho_S \partial^{\mu}\rho_S 
    + 2 m_B^2 \rho_S^2 - 2 m_B^2 J_V^{\mu} J_{V \mu}
  \nonumber\\
    && 
    -2  m_B J_V^{\mu}\partial^{\nu}({J_T}_{\nu \mu})\nonumber
    -\frac{1}{2}\partial^{\nu}({J_T}_{\nu \mu})
    \partial_{\sigma}({J_T}^{\sigma \mu}),
  \label{pos26}
  \end{eqnarray}
  \begin{eqnarray}
    \tilde{\mathcal L}_{S,Lx}^{PC(1)} 
    &=&  
    4 m_B \tilde{\alpha}_S\rho_S^3+2\tilde{\alpha}_S^2 \rho_S^4
    -4 m_B \tilde{\alpha}_S\rho_S J_V^{2} 
    +2\tilde{\alpha}_V\rho_S {J_T}_{\nu \mu}\partial^{\nu}(J_V^{\mu}) 
  \nonumber\\ 
  &&
  -2\tilde{\alpha}_S\rho_S J_V^{\mu}\partial^{\nu}({J_T}_{\nu \mu})
  - 2\tilde{\alpha}_S^2 \rho_S^2 J_V^{2}. 
  \label{pos27}
  \end{eqnarray}
 The densities and currents appearing in these formulas are special cases
  of the general ${\mathcal J}_i$ in Eq.(\ref{pis1}), they read
  \begin{eqnarray}
   \rho_{S} (x) & = & \sum_{\alpha} \bar{
   \Psi}_{\alpha} \Psi_{\alpha},~ ~ ~ ~ ~ ~ ~
    J_{V}^{  \mu}(x)  =  \sum_{\alpha}  \bar{
   \Psi}_{\alpha}\gamma^{\mu} \Psi_{\alpha},\nonumber\\
   \vec{\rho}_{tS}(x) & = & \sum_{\alpha}  \bar{\Psi}_{\alpha}\vec{\tau}
   \Psi_{\alpha}, ~ ~ ~ ~ ~ ~ 
    \vec{J}_{tV}^{  \mu}(x)  =  \sum_{\alpha}  \bar{
   \Psi}_{\alpha}\gamma^{\mu}\vec{\tau} \Psi_{\alpha},\nonumber\\
    J_{T}^{\nu  \mu}(x) & = & \sum_{\alpha}  \bar{
   \Psi}_{\alpha}\sigma^{\nu \mu} \Psi_{\alpha},~ ~ ~ ~ ~
    \vec{J}_{tT}^{\nu  \mu}(x)  =  \sum_{\alpha}  \bar{
   \Psi}_{\alpha}\sigma^{\nu \mu}\vec{\tau} \Psi_{\alpha}.
  \end{eqnarray}

  We see that  {\em the approximate form of
  the interaction effects in
  $ {\mathcal C}_{  }$ and $ {\mathcal B}_{  \mu}$ corresponds to
  having all possible terms made up of  products of different densities}. 
  
  Eq. (\ref{pos27})  define the scalar case of the
  approximation for the derivative exchange terms in
  Eq.~(\ref{pos13}).  
  The other cases can be derived similarly. The results
  are displayed in table \ref{tab00}.
\begin{table}
\centering
\begin{tabular}{|l|l|}\hline
Term & Expression\\
\hline
$\tilde{\mathcal
L}_{V, Lx}^{PC(0)}$ &  $-\frac{1}{2}\partial_{\mu} J_V^{\nu}\partial^{\mu} J_{V \nu} +2  m_B^2 J_V^{\nu}
J_{V \nu}$,\\
$\tilde{\mathcal L}_{V, Lx}^{PC(1)}$ & $ -2 \tilde{\alpha}_V^2 {J_{V}}^4 +4
m_B\tilde{\alpha}_S\rho_S J_V^{2}+ 2\tilde{\alpha}_S^2 \rho_S^2 J_V^{2}$ \nonumber\\&$2
\tilde{\alpha}_S \rho_S \partial^{\nu}(J_{T \nu \mu} J_V^{\mu}) \cdot$\\
$\tilde{\mathcal
L}_{T, Lx}^{PC(0)}$& $\frac{1}{2}\partial_{\mu} (J_T^{\alpha \beta})\partial^{\mu}
(J_{T \alpha \beta})$,\\
$\tilde{\mathcal L}_{T, Lx}^{PC(1)}$ &  $0$.\\
$\tilde{\mathcal L}_{tS, Lx}^{PC(0)}$ &$-\frac{1}{2}\partial_{\mu}\vec{\rho}_{tS}\partial^{\mu}\vec{\rho}_{tS}
+ 2 m_B^2 \vec{\rho}_{tS}^2 - 2 m_B^2 \vec{J}_{tV}^{2} $\nonumber\\ &$-2
m_B \vec{J}_{tV}^{\mu} \partial^{\nu} (\vec{J}_{tT \nu
\mu})-\frac{1}{2}\partial^{\nu}(\vec{J}_{tT \nu \mu})\partial_{\sigma}(\vec{J}_{tT}^{\sigma
\mu})$,\\
$\tilde{\mathcal L}_{tS, Lx}^{PC (1)}$ &  $-4 \tilde{\alpha}_S m_B \rho_S ( \vec{J}_{tV}^{2} -
\vec{\rho}_{tS}^2) - 2\tilde{\alpha}_S^2  \rho_S^2 (
\vec{J}_{tV}^{2} -
\vec{\rho}_{tS}^2)$\nonumber\\ &$-  2 \tilde{\alpha}_S\partial^{\sigma}(\vec{J}_{tT \sigma
\mu})   \rho_S
\vec{J}_{tV}^{\mu}+\tilde{\alpha}_V \vec{\rho}_{tS}\vec{J}_{tT \sigma
\mu}\partial^{\sigma}( J_{V}^{\mu})
\cdot$\\
$\tilde{\mathcal
L}_{tV, Lx}^{PC(0)}$ & $-\frac{1}{2}\partial_{\mu} \vec{J}_{tV}^{\nu}\partial^{\mu} \vec{J}_{tV \nu}
+2  m_B^2 \vec{J}_{tV}^{2}$, \\
$\tilde{\mathcal L}_{tV, Lx}^{PC(1)}$& $ 4 \tilde{\alpha}_S
m_B \vec{J}_{tV}^{2}  
\rho_S +2 \tilde{\alpha}_S^2 \rho_S^2 \vec{J}_{tV}^{2}$ \nonumber\\&$-
2 \tilde{\alpha}_V^2 {(\vec{J}_{tV}^{\nu} J_{V \nu})}^2-2 \tilde{\alpha}_S\rho_S
\partial^{\nu}(\vec{J}_{tT \nu \mu} \vec{J}_{tV}^{\mu}) \cdot$\\
$\tilde{\mathcal
L}_{tT, Lx}^{PC (0)}$ & $\frac{1}{2}\partial_{\mu} (\vec{J}_{tT}^{\alpha \beta})\partial^{\mu}
(\vec{J}_{tT \alpha \beta})$,\\
$\tilde{\mathcal L}_{tT, Lx}^{PC (1)}$ & $0$.\\
\hline
\end{tabular}\\
\caption{The different terms (except the scalar term) of the approximate Lagrange density.\label{tab00}}
\end{table}
Inserting these results into ${\mathcal L}
_{\rm L  ~ eff}^{PC}$ = $\langle\phi_0|: \hat{\mathcal L}
_{L}^{PC}:|\phi_0 \rangle$ yields:
\begin{eqnarray}
  {\mathcal L}_{\rm L ~ eff}^{PC} 
  &=& 
  \sum_{\alpha=A} \bar{\Psi}_{\alpha} ( i \gamma_{\mu} \partial^{\mu} -m_B ) 
  \Psi_{\alpha} 
\nonumber\\  
 &&  
  -\frac{1}{2} \tilde{\alpha}_S \rho_S^2 
  -  \frac{1}{2} \tilde{\alpha}_{tS} \vec{\rho}_{tS}^2 
  - \frac{1}{2} \tilde{\alpha}_VJ_V^{2} 
  - \frac{1}{2} \tilde{\alpha}_{tV}  \vec{J}_{tV}^{2}  
\nonumber\\  
  && 
  -\frac{1}{2}\tilde{\delta}_S\partial_{\mu}\rho_S\partial^{\mu}\rho_S 
  - \frac{1}{2} \tilde{\delta}_{tS}\partial_{\mu}\vec{\rho}_{tS}
    \partial^{\mu}\vec{\rho}_{tS} 
  - \frac{1}{2}\tilde{\delta}_V \partial_{\nu}J_V^{\mu}
    \partial^{\nu}J_{V \mu}
\nonumber\\ 
  && 
  -\tilde{\delta}_{tV}\partial_{\nu}\vec{J}_{tV}^{\mu}
    \partial^{\nu}\vec{J}_{tV\mu} 
  -  \frac{1}{2} \tilde{\theta}_TJ_V^{\mu}\partial^{\nu}({J_T}_{\nu\mu})
  - \frac{1}{2} \tilde{\theta}_{tT}\vec{J}_{tV}^{\mu}
    \partial^{\nu}(\vec{J}_{tT \nu \mu}) 
\nonumber\\
  && 
  +{\mathcal L}_{Lx}^{PC(1)}+ {\mathcal L}_T^{corr} + {\mathcal L}_{ A},
\\
 {\mathcal L}_{Lx}^{PC(1)}
 &=&
 c_S \tilde{\mathcal L}_{S,Lx}^{PC(1)}
 +c_V \tilde{\mathcal L}_{V,Lx}^{PC(1)}
 +c_{tS} \tilde{\mathcal L}_{tS,Lx}^{PC(1)}
 +c_{tV} \tilde{\mathcal L}_{tV,Lx}^{PC(1)}
\\
  {\mathcal L}_T^{corr} 
  &=& 
  \frac{1}{16}\delta_S (\frac{1}{2}
        \partial_{\mu} (J_T^{\alpha \beta})\partial^{\mu}(J_{T \alpha \beta})
   +\frac{1}{2}\partial_{\mu} (\vec{J}_{tT}^{\alpha \beta})\partial^{\mu}
   (\vec{J}_{tT \alpha \beta}))
\nonumber\\ 
  &&
  + \frac{1}{16}\alpha_S (J_{T \nu \mu}J_{T}^{\nu \mu} +
  \vec{J}_{tT \nu \mu} \vec{J}_{tT}^{\nu \mu}) 
\nonumber\\  
  &&
  + \frac{1}{2}[c_S
  \partial^{\nu} (J_{T \nu \mu})\partial_{\sigma}(J_{T}^{\sigma \mu})
  + c_{tS}\partial^{\nu}(\vec{J}_{tT \nu \mu})
   \partial_{\sigma}(\vec{J}_{tT}^{\sigma \mu}) ],
\end{eqnarray}
with primed coupling constants $\tilde\alpha_i$, $\tilde\delta_i$,
$\tilde\theta_i$. These are not new parameters but can be related
uniquely by a linear transformation
to the given original couping parameters of the model
$$
  (\tilde\alpha_S,
  \tilde\alpha_{tS},
  \tilde\alpha_V,
  \tilde\alpha_{tV},
  \tilde\delta_S,
  \tilde\delta_{tS},
  \tilde\delta_V,
  \tilde\delta_{tV},
  \tilde\theta_T,
  \tilde\theta_{tV})
  \quad\Leftarrow\quad
  (\alpha_S,
  \alpha_V,
  \alpha_{tV},
  \delta_S,
  \delta_V,
  \delta_{tV})
$$
The actual transformation is a bit lengthy. It is given explicitely
in appendix \ref{Anhang:par}.

Finally,  ${\mathcal L}_{ A}$ is an electromagnetic Lagrangian density
with an exchange correction, it just remains to approximate the
exchange part  by the  local
density (Slater) approximation~\cite{ben}.
We have seen that in the Gordon decomposition
representation, the dominant part of the ${\mathcal B}_{j \mu}$ and
${\mathcal C}_{ j }$ densities can be replaced effectively by
tensor terms and all possible mixings or combinations of nonlinear terms
in all orders. It means that if we perform Hartree calculations by
taking into account tensor terms and all possible mixings or combinations of nonlinear terms
of all orders, the effect of the dominant part of ${\mathcal B}_{ j \mu}$ and
${\mathcal C}_{ j }$ is already represented effectively, but still
a part which manifests itself as an effect beyond the Dirac equation  and
retardation effects are not yet represented. A strict Hartree theory
including these additional effects can be constructed by starting off
with the appropriate terms, but at the cost of introducing new
parameters (the details about the mapping from Hartree-Fock to
Hartree can be seen in Appendix (3)).

The new terms that emerge in our investigation lead to modifications of a) the spin-orbit force due to tensor terms, b) the isovector channel due to products of isoscalar and isovector terms and the inclusion of isovector-scalar terms and c) the density dependence of the model. They in turn lead to a different density dependence of the effective mass. A similar change of the spin-orbit force and the effective mass might allow a larger effective mass closer to the values predicted by Skyrme forces. It is interesting to discover the effect of the modified density dependence in both the isoscalar and the isovector channel.
All these contributions arise without introducing new parameters.  

This unavoidably complicated mapping of the original exchange terms
into a manageable Lagrangian in this rough approximation will now be
applied to typical nuclear systems.

\section{Finite Nuclei}
Since the model discussed in this paper is meant to be applied to finite nuclei, observables connected with them should be considered when evaluating the consequences of our approximate treatment of exchange terms. We will take a look at a variety of different observables. The calculations are performed in spherical symmetry.

As a preliminary step, we will apply further approximations.
Let us take the following Lagrange density: 
\begin{eqnarray}
  {\mathcal L} _{\rm NL ~ eff}^{PC} 
  &=& 
  {\mathcal L} _{\rm free}
  + \tilde{\mathcal L}_{\rm L}^{PC}
  +{\mathcal L}_{\rm Lx ~ eff}^{PC(0)}
  + {\mathcal L}_{\rm NL}^{PC } 
  +{\mathcal L} _{\rm A} 
  +{\mathcal L} _{\rm Ax} ,
\label{dudu}
\end{eqnarray}
with 
\begin{eqnarray}
  {\mathcal L} _{\rm free}
  &=& 
  \sum_{\alpha=A} \bar{\Psi}_{\alpha} ( i \gamma_{\mu} \partial^{\mu} -m_B ) 
  \Psi_{\alpha},
\nonumber\\  
  \tilde{\mathcal L}_{\rm L}^{PC}
  &=&
  -\frac{1}{2}  \tilde{\alpha}_S \rho_S^2 - \frac{1}{2}  \tilde{\alpha}_V
    J_V^{2} - \frac{1}{2}  \tilde{\alpha}_{tV}
   \vec{J}_{tV}^{2}  - \frac{1}{2}  \tilde{\delta}_S
  \partial_{\mu}\rho_S\partial^{\mu}\rho_S 
\nonumber\\
  && 
  -\frac{1}{2}\tilde{\delta}_V \partial_{\nu}J_V^{\mu}\partial^{\nu}J_{V \mu} 
  -\frac{1}{2}\tilde{\delta}_{tV}\partial_{\nu}\vec{J}_{tV}^{\mu}
   \partial^{\nu}\vec{J}_{tV \mu},
\nonumber\\ 
  {\mathcal L}_{\rm Lx ~ eff}^{PC(0)}
  &=& 
  -\frac{1}{2}  \tilde{\alpha}_{tS} \vec{\rho}_{tS}^2  
  - \frac{1}{2}  \tilde{\delta}_{tS}
  \partial_{\mu}\vec{\rho}_{tS}\partial^{\mu}\vec{\rho}_{tS}
  -  \frac{1}{2}  \tilde{\theta}_TJ_V^{\mu}\partial^{\nu}({J_T}_{\nu\mu}) 
  - \frac{1}{2}  \tilde{\theta}_{tT}
  \vec{J}_{tV}^{\mu}\partial^{\nu}(\vec{J}_{tT \nu \mu}),
\nonumber\\
  {\mathcal L}_{\rm A}
  &=& 
  -\frac{1}{2}\partial_{\nu}A^{\mu}\partial^{\nu}A_{\mu}
  + e \sum_{\alpha} \bar{ \Psi}_{\alpha}
    \gamma^{\mu}A_{\mu} \frac{1}{2}(1+\tau_3)\Psi_{\alpha},
\nonumber
\end{eqnarray}
and 
\begin{eqnarray}
 {\mathcal
L}_{\rm NL}^{PC}=-\frac{1}{3}\beta_S \rho_S^3 -\frac{1}{4}\gamma_S \rho_S^4. \nonumber
\end{eqnarray}
Thus now $ {\mathcal L}_{Lx}^{PC(1)}$,  ${\mathcal
L}_T^{corr}$, and ${\mathcal
L}_{\rm NLx}^{PC }$ in Eq.~(\ref{dudu}) are neglected. ${\mathcal
L}_T^{corr}$ can be considered to give a small contribution, and it is
assumed that parts
of ${\mathcal L}_{Lx}^{PC(1)}$ and  ${\mathcal
L}_{\rm NLx}^{PC }$ are already absorbed effectively in ${\mathcal
L}_{\rm NL}^{PC}$, of course, this assumption is not really true in the
strict sense. 
We will study the following Lagrange densities for which we will
determine parameter sets as described in the next
section:
\begin{enumerate}
\item LH: Point-coupling model, no nonlinear terms, exchange
omitted
\begin{eqnarray}
{\mathcal L} _{\rm LH}&=& {\mathcal L} _{\rm free}
+ \tilde{\mathcal L} _{\rm L}^{PC}+{\mathcal L}_{\rm A}. \label{ps1}
\end{eqnarray}
\item LHx : Point-coupling model, no nonlinear terms, approximate exchange
included
\begin{eqnarray}
{\mathcal L} _{\rm LHx}= {\mathcal L} _{\rm LH}+{\mathcal L}
_{\rm Lx ~ eff}^{PC(0)}+{\mathcal L} _{\rm Ax}.
\end{eqnarray}
\item NLH   : Point-coupling model,  nonlinear terms, exchange omitted
\begin{eqnarray}
{\mathcal L} _{\rm NLH}= {\mathcal L} _{\rm LH}+{\mathcal L}
_{\rm NL}^{PC}.
\end{eqnarray}
\item NLHx  : Point-coupling model,  nonlinear terms, nonlinear
exchange omitted but linear exchange terms included
\begin{eqnarray}
{\mathcal L} _{\rm NLHx}={\mathcal L} _{\rm LHx}+{\mathcal L}
_{\rm NL}^{PC}.
\label{ps4}
\end{eqnarray}
\end{enumerate}
Comparing these four versions should shed light on both the role of
nonlinear terms and of exchange for realistic applications.

\subsection{Determination of Parameters}
We follow
the $\chi^2$ fitting procedure of Refs. \cite{pg,rufa}, choosing the same 
set of physical observables (binding energies, diffraction radii and
surface thicknesses of $ ^{16}O, ^{40}Ca,^{48}Ca, ^{58}Ni,
^{90}Zr,^{116}Sn, ^{124}Sn $, and $^{208}Pb$ ) and weights (0.2$\%$ relative
error for the binding energies, 0.5 $\%$ for the diffraction radii and 1.5$\%$ for
the surface thicknesses)
using the constant gap pairing correlation~\cite{pg,rufa}. Our
parameters are $\bar{\alpha}_S$, $\bar{\alpha}_V$, $\bar{\alpha}_{tV}$, $\bar{\delta}_S$, $\bar{\delta}_V$ and
$\bar{\delta}_{tV}$ (six
parameters) for LHx and LH and additionally $\beta_S$ and $\gamma_S$ for
NLH and NLHx.

\subsection{Results}

\begin{table}[h]
\centering
\begin{tabular}{|c|c|c|c|c|} \hline 
Model &  $ \chi^2_{\rm E_B}$ & $\chi^2_{\rm R_d}$  &
$\chi^2_{\sigma}$ &  $ \chi^2_{\rm Total}$ \\\hline \hline
LH & 872.21 & 70.09 & 1250.96 & 2193.27\\
LHx &89.31&18.58& 429.31&537.20\\
NLH & 27.65 & 6.57 & 75.79 & 110.01\\
NLHx & 21.93 & 6.91 & 59.51 & 88.35\\\hline
\end{tabular}\\
\caption{$ \chi^2$ results from the models defined by Eqs. (\ref{ps1}-\ref{ps4}).
$\rm E_{B}$ denotes the binding energies, $\rm R_{d}$  the diffraction
radii and $\sigma $ the surface thicknesses\label{tab3}}
\end{table}
\begin{table}[h] 
\centering
\begin{tabular}{|c|c|c|c|c|c|c|c|c|}\hline
Model & ${\alpha}_S$ & ${\alpha}_V$ & ${\alpha}_{tV}$ &
${\delta_S}$ &  ${\delta}_V$ & ${\delta}_{tV}$  & $\beta_S$ &  $\gamma_S$  \\\hline \hline
LHx & 27.632 & -14.881 & 27.623& -0.517& -0.152& 0.183 & 0 & 0  \\
LH & -16.706 & 12.829 & 2.152& -1.712& 0.670 &0.100 & 0 & 0\\
NLHx & -12.315 & 2.193 & 3.110& -0.721& -0.041 &0.045 & 24.168 & -83.312\\
NLH & -14.726 & 9.887 & 1.345& -0.746& -0.058 & -0.377 &23.734 & -81.844  \\\hline
\end{tabular}\\
\caption {Coupling constants of the various models\label{tab4}}
\end{table}

Table~\ref{tab3} collects the $\chi^2$ results from the four model variations. Here LHx, LH, NLHx, NLH denote parameter sets
from Eqs. (\ref{ps1}-\ref{ps4}). $E_B$, $R_d$ and $\sigma$ denote the binding
energy, diffraction radius, and surface thickness.

\begin{figure}[htb]
\centering
\subfigure{\epsfig{figure=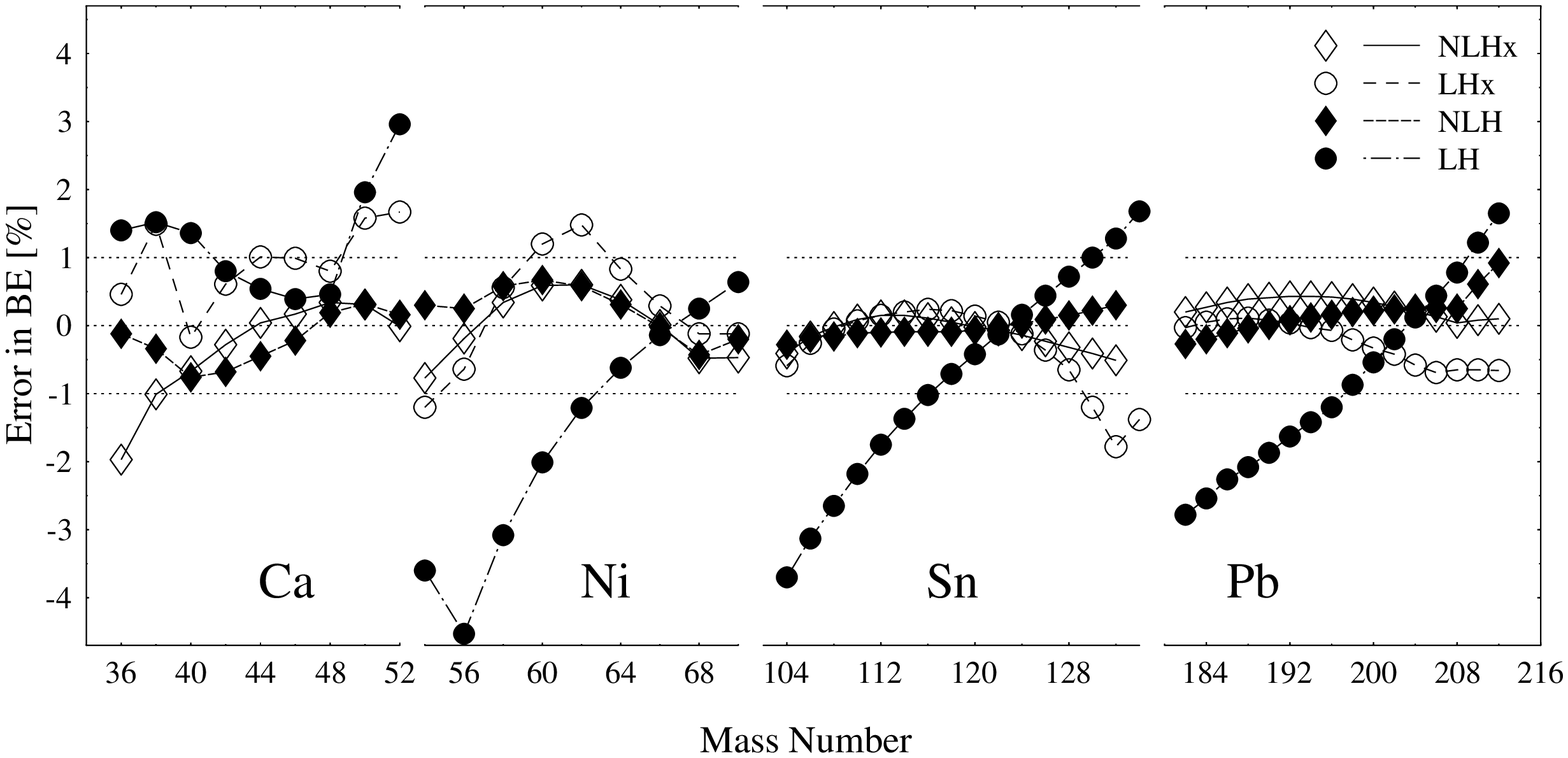,width=.8\textwidth}}
\centering
\subfigure{\epsfig{figure=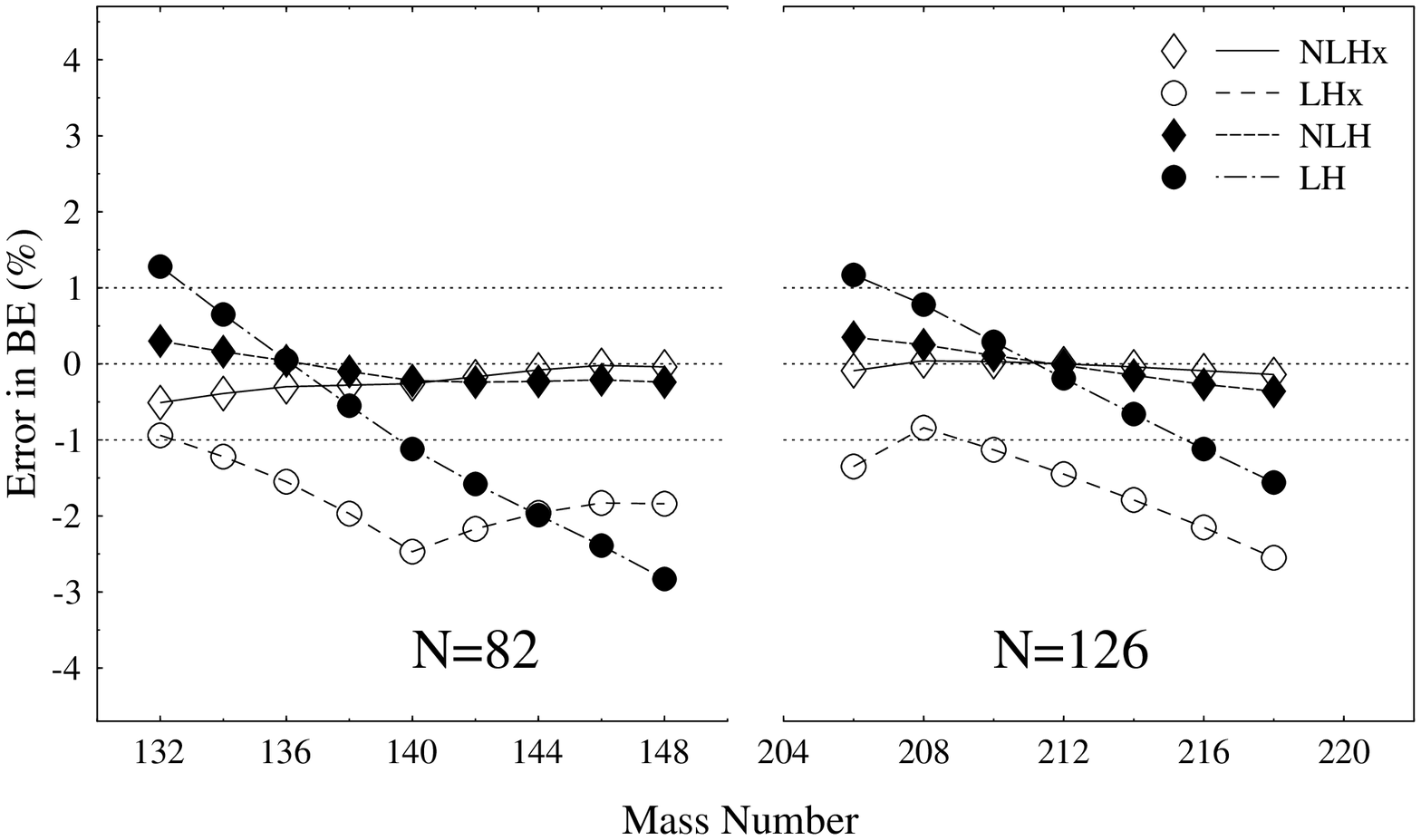,width=.8\textwidth}} 
\caption{Error in the binding energies for selected isotopes (upper part) and isotones (lower part).}\label{gam1}
\end{figure}

To see the difference among the four parameter sets clearly in
binding energies of finite nuclei, we use the error
in the binding energies ($ {\mathcal E}_{\rm E_B}$) and the two-neutron ($S_{2n}$) and the two-proton
($S_{2p}$) separation energies from some isotopic and isotonic chains. 
The error in the binding energy is defined as
\begin{eqnarray}
  {\mathcal E}_{\rm E_B} = \frac{\rm E_{B ~ th} - E_{ B ~ exp}}{ \rm E_{ B
~ exp}} ,    
\end{eqnarray}
while $S_{2p}$ and  the $S_{2n}$ are defined as
\begin{eqnarray}
 S_{2n} &=& {\rm E_B}(N,Z) - {\rm E_B}(N-2,Z),\nonumber \\
 S_{2p} &=& {\rm E_B}(N,Z) - {\rm E_B}(N,Z-2)
\end{eqnarray}   
respectively, with ${\rm E_B}$(N,Z) being the calculated binding
energy. 
Fig.~\ref{gam1} shows that in the isotope chains, LHx has
better $ {\mathcal E}_{\rm E_B}$ results than LH, but in the isotonic
chains, LH performs better. Thus even though LHx has a better $\chi_{E_B}^2$ than LH, it does not mean it must have better binding
energy predictions for all spherical nuclei. Although the
$\chi^2_{E_B}$ of NLH and NLHx are not drastically different from
that of LHx, they yield much better predictions than LHx for these
isotopic and isotonic trends.We see here that the role of nonlinear
terms for an acceptable $ {\mathcal E}_{\rm E_B}$ prediction is more important than
that of the tensor terms, which in this case partly represents the
exchange effect of the linear point-coupling model. NLH has a quality in ${\mathcal E}_{\rm E_B}$ as good as NLHx.  
\begin{figure}[htb]
\centering
\mbox{\subfigure{\epsfig{figure=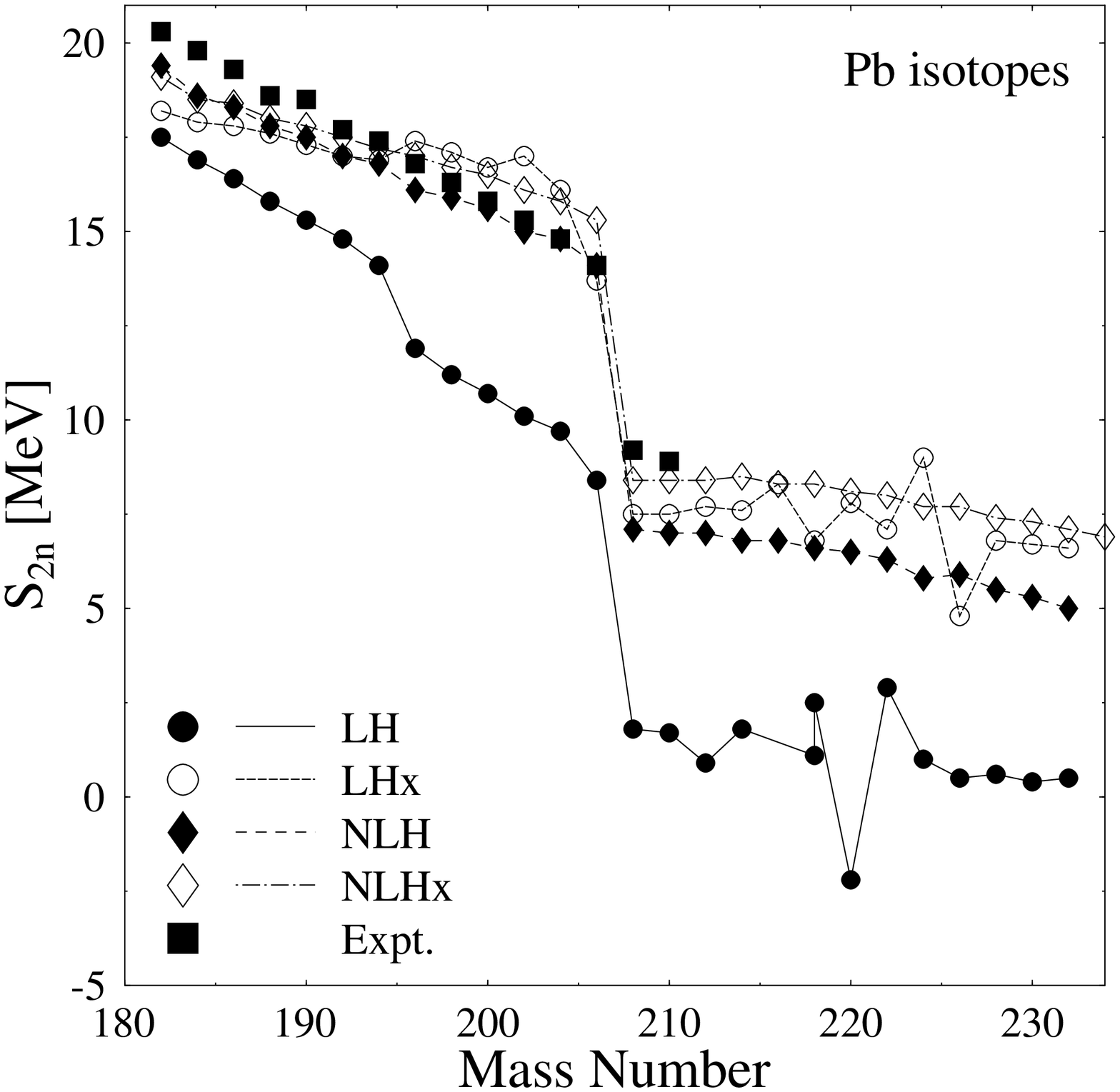,width=.5\textwidth}}\quad
      \subfigure{\epsfig{figure=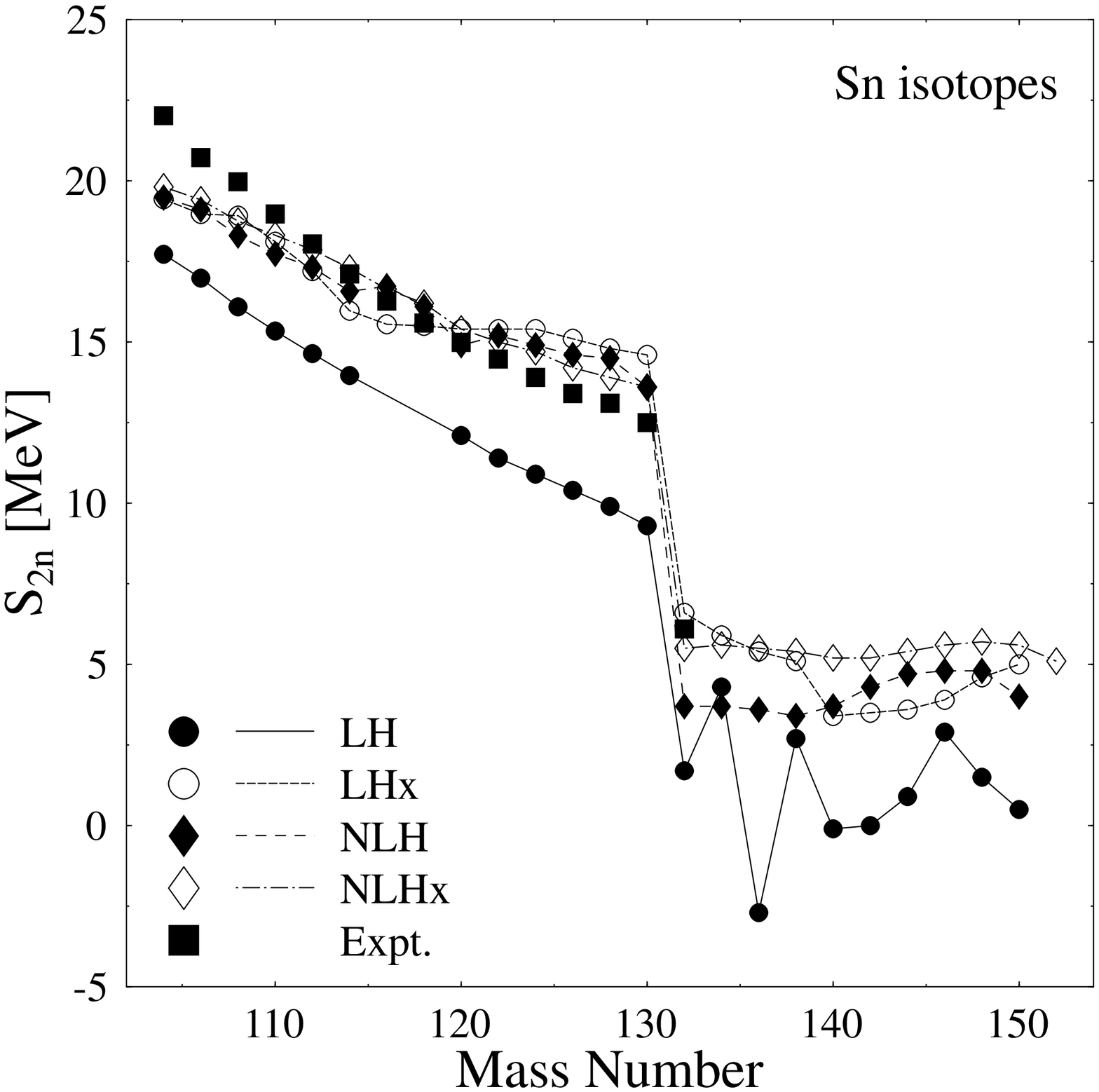,width=.5\textwidth}} }
\centering
\mbox{\subfigure{\epsfig{figure=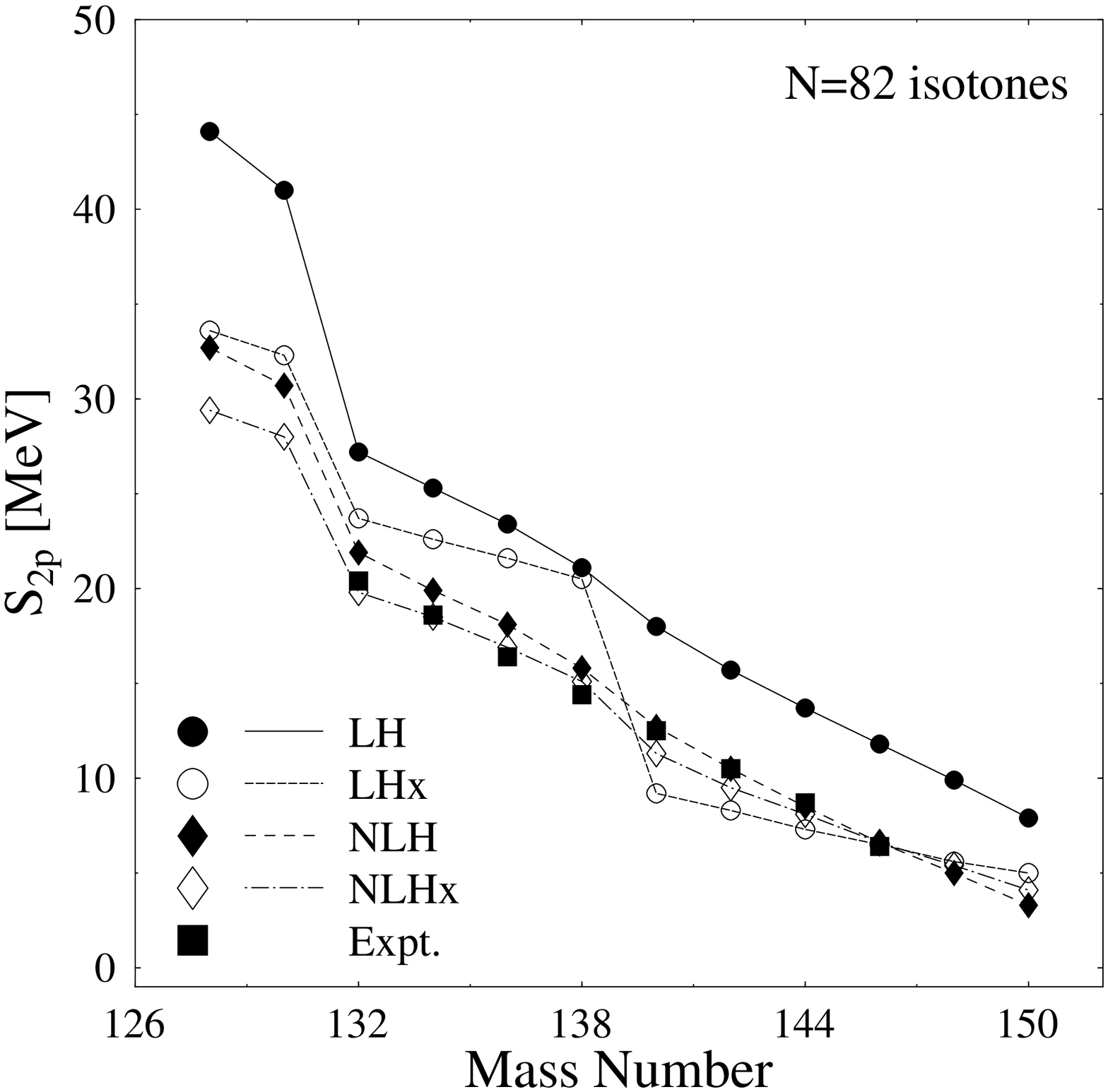,width=.5\textwidth}}\quad
      \subfigure{\epsfig{figure=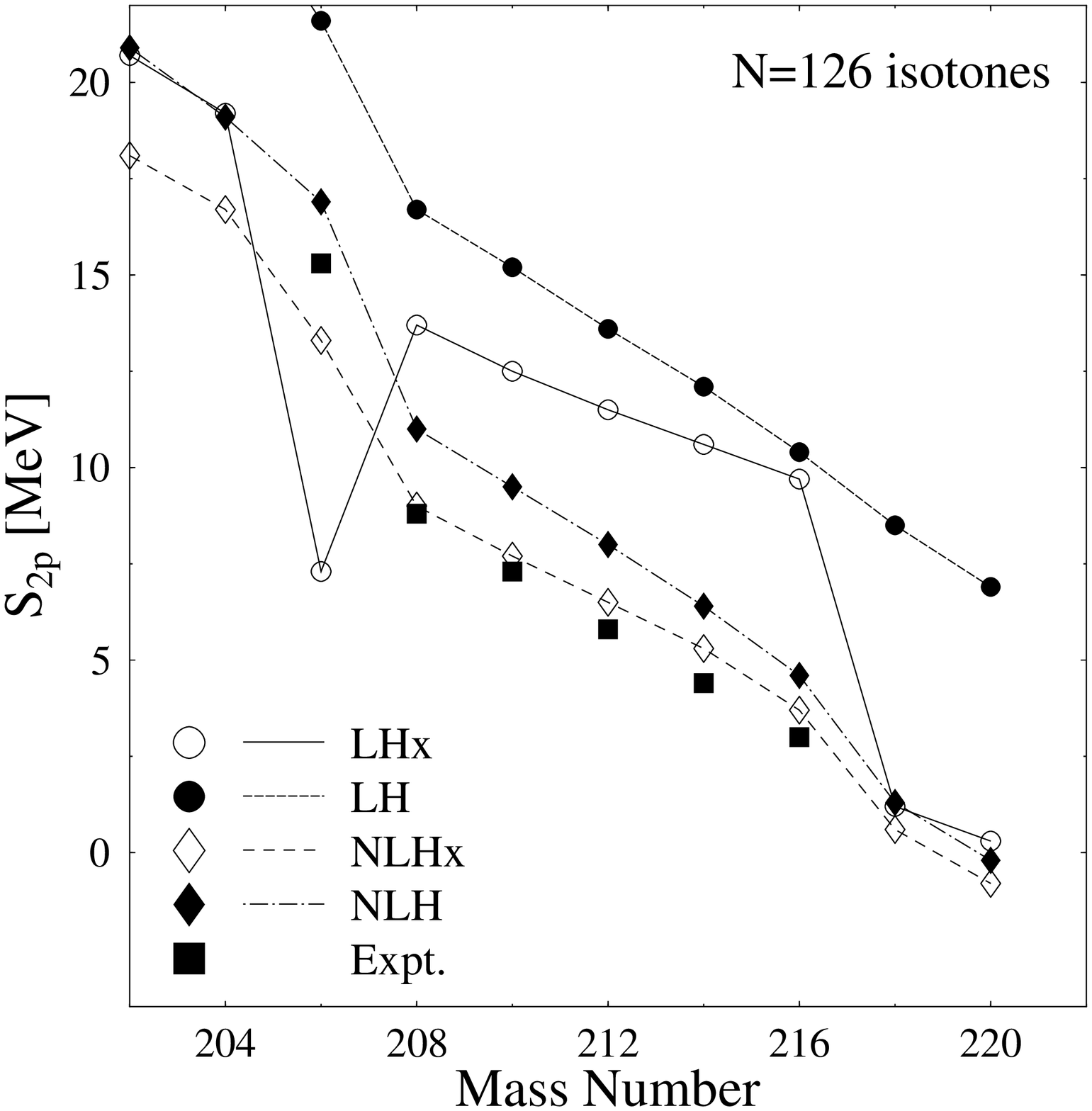,width=.5\textwidth}} }
\caption{Two-neutron separation energies ($S_{2n}$) for lead and tin isotopes
(top)
and two-proton separation energies ($S_{2n}$) for N=82 and N=126 isotone chains N=82 (bottom).}\label{gam2}
\end{figure}

Fig.~\ref{gam2} shows that on the average, $S_{2n}$ and $S_{2p}$ from LHx are better than the ones from LH. Here we can see 
an important role of the nonlinear terms for the prediction of  binding energies, because only 
NLH and NLHx can reproduce the $S_{2n}$ and $S_{2p}$ experimental data for
almost all represented isotopes and isotones. The NLHx results are
closer to the experimental data than NLH. The effect of the presence
of tensor terms in the nonlinear model cannot be seen from $ {\mathcal E}_{\rm E_B}$, but
$S_{2n}$ and $S_{2p}$ show
better trends. 

\begin{figure}[htb]
\centering
\mbox{\subfigure{\epsfig{figure=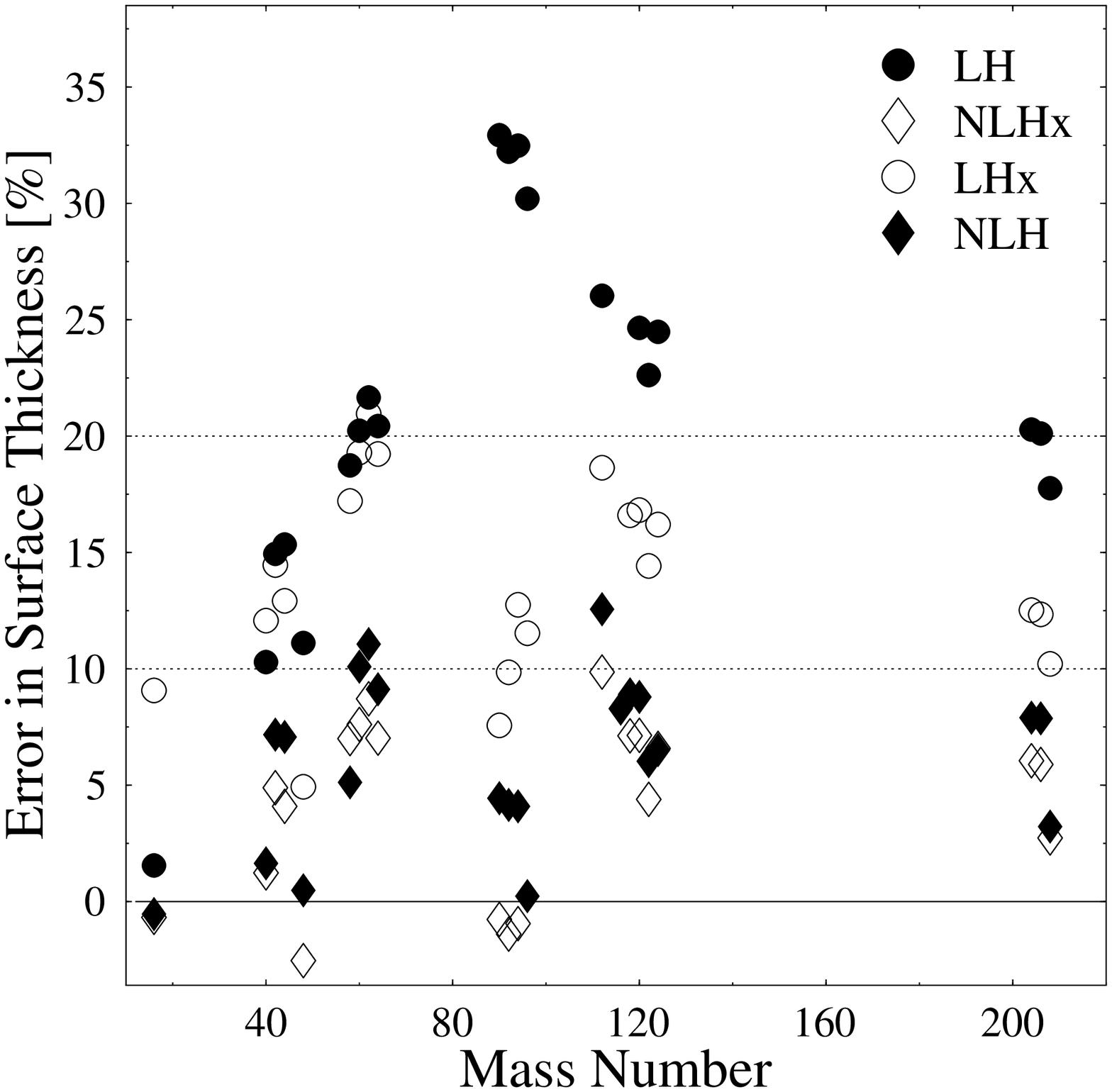,width=.5\textwidth}}\quad
      \subfigure{\epsfig{figure=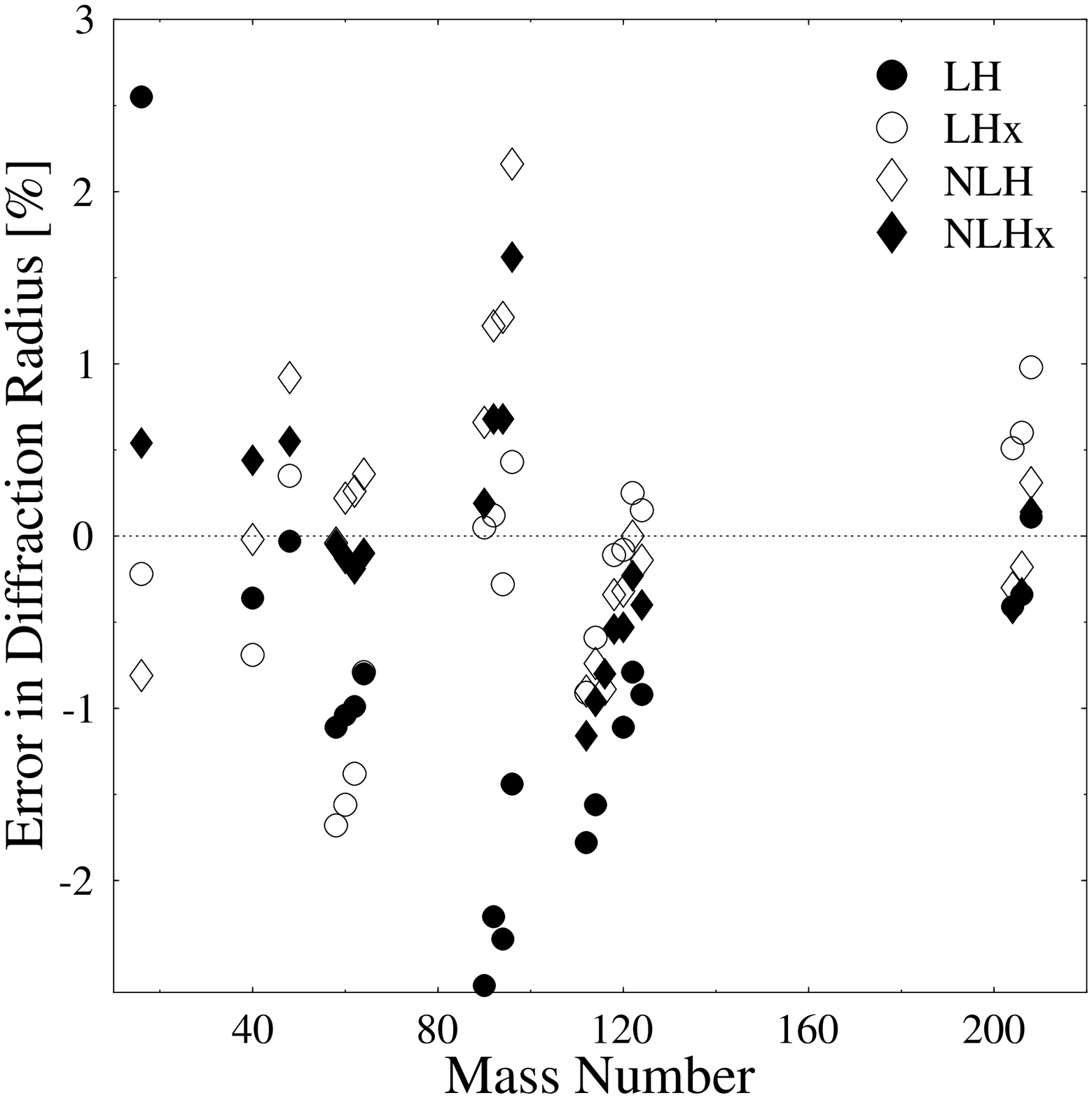,width=.5\textwidth}} }
\caption{Error in the surface thicknesses and diffraction radii for
some selected isotopes.}\label{gam3}
\end{figure}
\begin{figure}[htb]
\centering
\mbox{\subfigure{\epsfig{figure=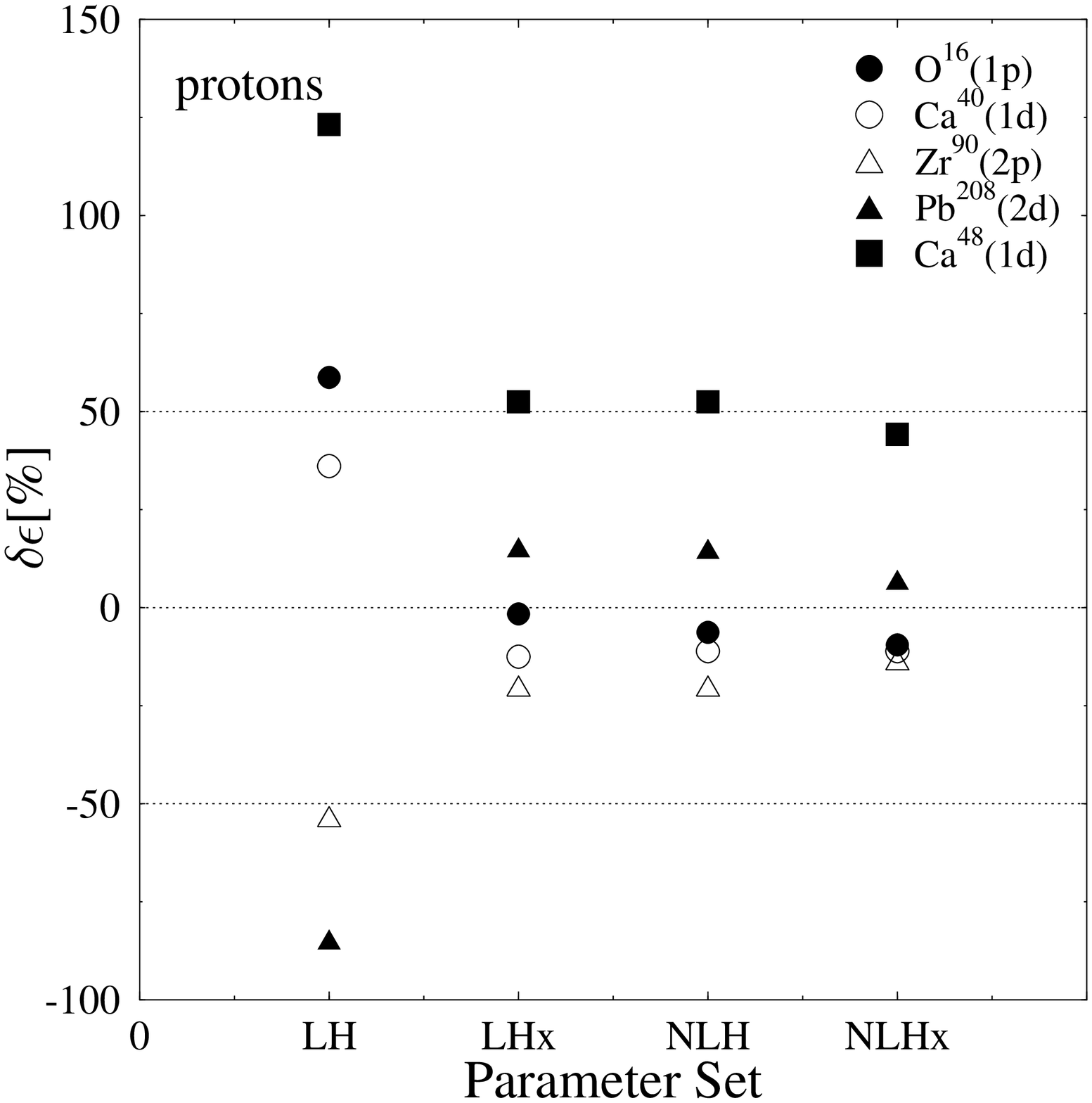,width=.5\textwidth}}\quad
      \subfigure{\epsfig{figure=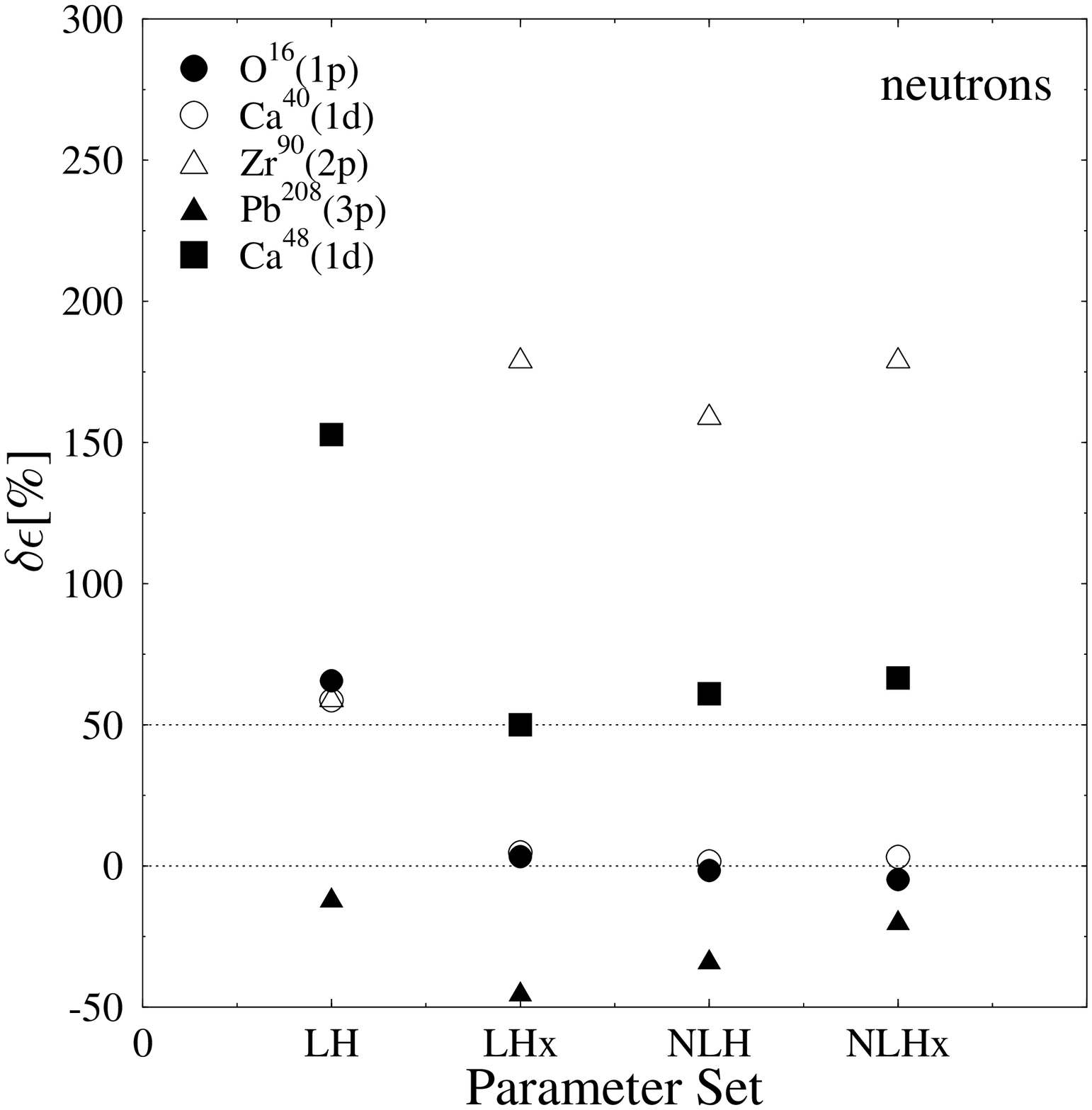,width=.5\textwidth}} }
\centering
\mbox{\subfigure{\epsfig{figure=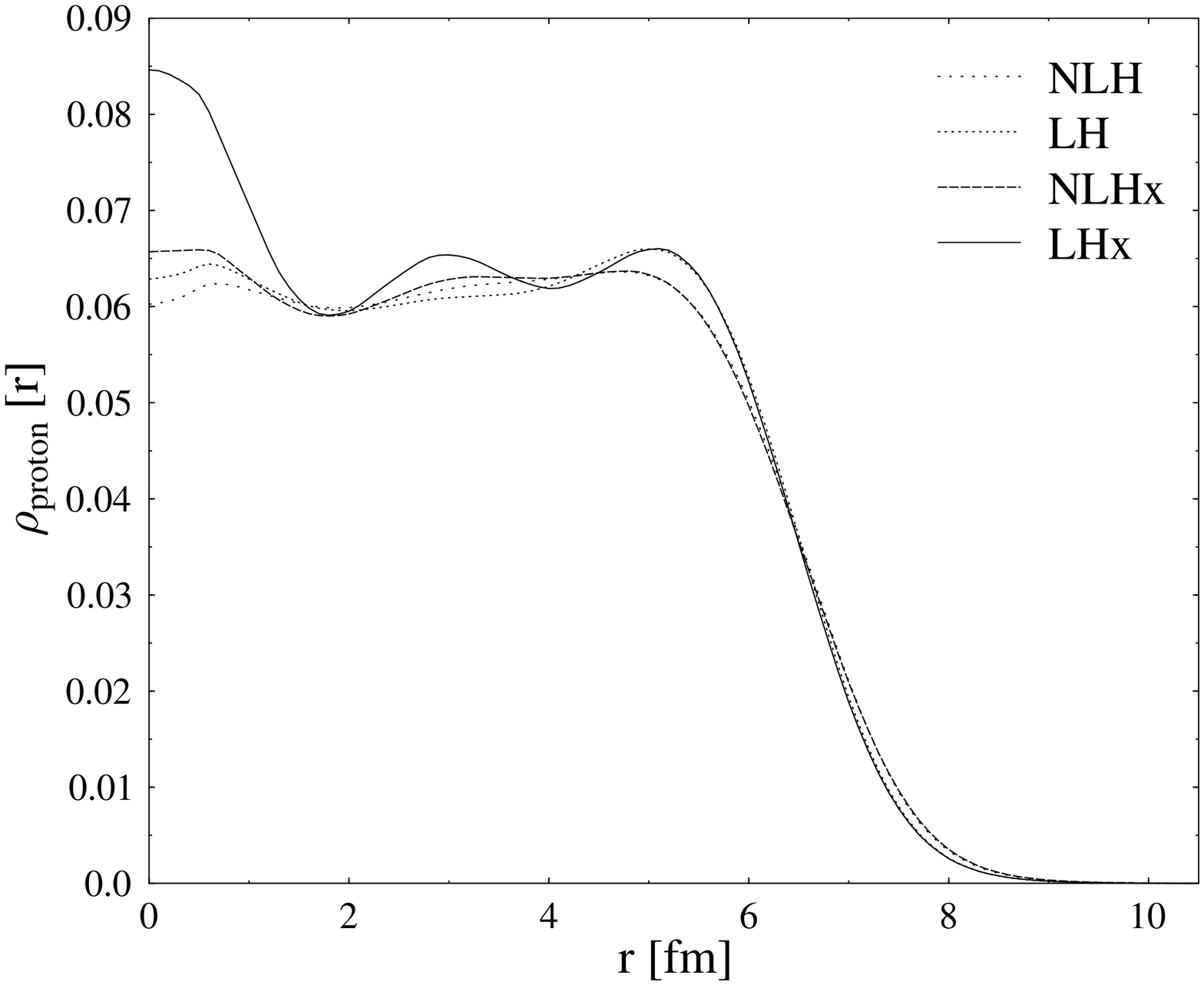,width=.5\textwidth}}\quad
      \subfigure{\epsfig{figure=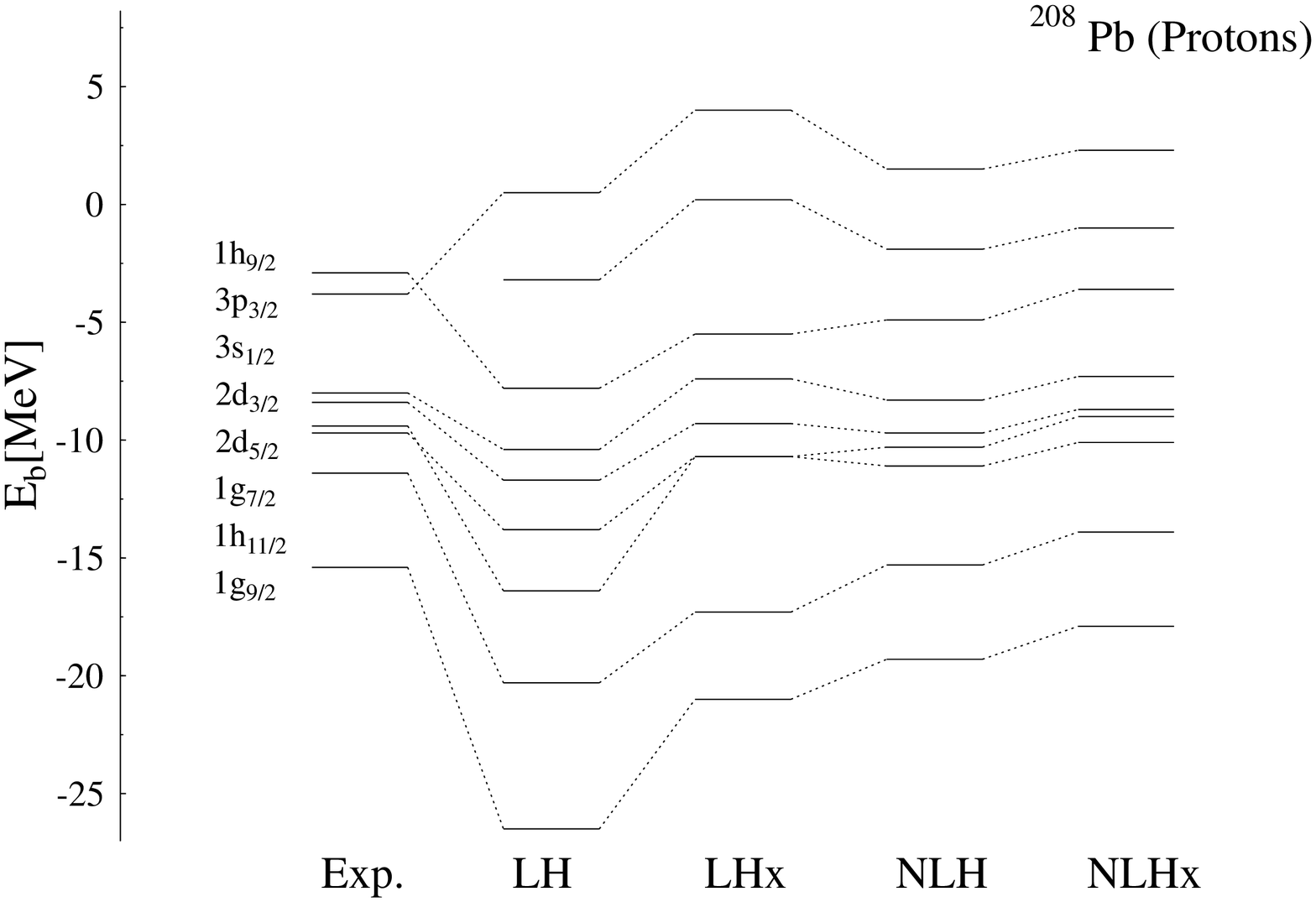,width=.5\textwidth}} }
\caption{Error in the spin-orbit splittings for some spherical nuclei
(top) and the radial proton density and proton single-particle
spectrum for $^{208}$Pb (bottom). }
\end{figure}

Errors in the diffraction radii ($ {\mathcal E}_{\rm R_d}$)and the surface
thicknesses ($ {\mathcal E}_{\sigma}$) are defined similarly as for the binding
energies. The left picture of figure~\ref{gam3} shows that LH has an
 $ {\mathcal E}_{\sigma}$ of more than 20 $\%$ . LHx has an $ {\mathcal E}_{\sigma}$  value in the range
below 20 $\%$. It is still a bad result compared to NLHx and NLH
which have an  $ {\mathcal E}_{\sigma}$ value below 10 $\%$.  Although the
$\chi^2_{\sigma}$ of NLHx is considerably better than for NLH, they yield similar
predictions for the surface thicknesses. This confirms the result of Ref. \cite{nley}, namely
that the relativistic linear Hartree-Fock calculation still cannot give acceptable surface
thickness predictions. Here it is clearly shown that the exchange effect 
has an important role in the surface thickness prediction. This
figure also shows that only models which include the nonlinearities
can give acceptable predictions.  The right picture
of figure~\ref{gam3} shows that the  $ {\mathcal
E}_{\rm R_d}$ results of the parameter sets
behave quite differently in each isotopic chain, but as the difference is not really significant, 
they may basically be considered to yield similar predictions in  $ {\mathcal
E}_{\rm R_d}$.

The definition of the error in spin-orbit 
splitting is the same as in the previous cases. For the neutron case, all
models give a bad prediction for $^{90}Zr (2p)$ and $^{48}Ca (1d)$. Only LH gives acceptable results in $^{208}Pb (3p)$, LHx
and NLH in $^{16}O (1p)$ and $^{40}Ca (1d)$, and NLHx in
$^{208}Pb (3p)$, $^{16}O (1p)$ and $^{40}Ca (1d)$. For protons,
LH cannot give acceptable results. LHx and NLH have
three acceptable results while NLHx has four, all
parameter sets cannot reproduce the experimental result of
$^{48}Ca$(1d) for protons.
LHx is intermediate in
quality between LH, on the one hand, and NLH and NLHx on the other
for the single-particle spectra. From these facts it is clear that the
exchange effects from the linear terms are important for the shell structure
though quantitatively still insufficient, while the nonlinearity in the
relativistic mean-field model clearly appears necessary for shell
structure predictions. For the case of $^{208}$Pb, LHx shows 
strong fluctuations in the surface part of the density and yields quite a
large value of $\rho_P$(r) in the center. This fact is probably due
to the rough approximation used for the exchange of
the derivative terms. We can see the role of the nonlinearity in
NLHx which remedies the central density in comparison to LHx.

The above results can be understood from the nonrelativistic
analysis~\cite{otna} as follows: the nonlinear models (NLH, NLHx)
exhibit better predictions for finite nuclei observables than the linear
ones (LH,LHx) because they have an adequat density dependence in
the central and spin-orbit potentials. LHx shows better predictions than
LHx in shell structure related observables because LHx contains a density dependence in the spin-orbit potential. LHx has better predictions than LH in binding energy and
surface thickness, but the result is still far from acceptable
(in contrast to NLH, NLHx) because LHx gets the density dependence in the central potential not from the dominant part $C_1$ but from the minor part $C_6$.    

In summary, it is found that the exchange effect in the linear
Hartree-Fock point-coupling model cannot be absorbed effectively into
the coupling
constants of a Hartree calculation. In agreement with Refs.~\cite{nley,schmd}
where the linear Walecka model was used, a similar situation happens in
the linear point-coupling model, namely that Hartree and Hartree-Fock
calculations cannot give predictions which are close to experimental
data. The nonlinear terms are needed to remedy this. Upon introducing
the nonlinear terms, the exchange effect from the linear terms does
not drastically show up. Thus it seems that the nonlinear terms are
more important than the exchange effect from the linear terms. The validity of
this presumption still needs to be checked, however, by the
exact calculation, namely taking into account the exchange of both the linear and
nonlinear terms exactly. 

\section{Conclusion}
We have shown that formally, a Lagrangian
densitiy ${\mathcal L}^{HF}$ of the standard ansatz can
be determined in a relativistic point-coupling Hartree-Fock sense that is not equivalent
to that determined in a relativistic point-coupling strict Hartree
sense, ${\mathcal L}^{H}$ due to the exchange of the linear derivative terms
and exchange of the nonlinear terms. The equivalency can be obtained if
we use a ``complete'' ansatz, but the cost we must pay is introducing
more parameters. The exchange of the linear
derivative terms is created by the densities $ {\mathcal C}_{i  }$ and $ {\mathcal
B}_{i 
\mu}$. By using the Gordon decomposition, we can separate these
densities into two parts. The role of the first part can be replaced effectively
by tensor terms and all possible nonlinear terms, while the second
part is a genuine feature of these densities. This second part is responsible for
creating effects beyond the Dirac equation, nonlocal and  retardation effects. An order of magnitude
analysis of the $ {\mathcal C}_{i  }$ and $ {\mathcal B}_{i 
\mu}$ densities in the Gordon decomposition representation of these
densities shows that these parts are expected to be small, yet the actual role of
these parts in finite nuclei observables is still interesting to
investigate. Also, as discussed before, taking into account the nonlinear exchange terms might alter the
picture significantly. 

Our study, by using rough approximations in the Lagrangian densities,
has shown gradual improvements in prediction for the finite nuclei. These results and the possibility that the effect of the exchange of the
nonlinear terms and some neglected effects in the approximate forms of
the  densities
$ {\mathcal C}_{i  }$ and $ {\mathcal B}_{i 
\mu}$, which could reveal more improvements or even give
a different picture  in finite
nuclei, support the importance of exact
Hartree-Fock point-coupling calculations.

Another interesting feature of such Hartree-Fock point-coupling calculations 
is the possibility to take into account terms which cannot appear in 
the Hartree calculation, for example the pionic degrees of freedom will contribute
via the exchange terms. 

A reexamination of the ansatz is needed to improve the effectivity of the model
and  a careful optimization procedure will be
mandatory for fitting the model, both need elaborate and extensive work. For these
reasons, we postpone the study of the complete quantitative Hartree-Fock model to future work. 
\begin{appendix}
\section{Appendix}
\subsection{From the Walecka to the Point-Coupling Model}

The advantages and disadvantages of
both models in the context of the Hartree-Fock calculation are discussed.
A Lagrangian density operator
for the Walecka (W) model is chosen as in~\cite{Lee}:
\begin{eqnarray}
\hat{{\mathcal{L}}}^{W}=\hat{{\mathcal{L}}} _{L}^{W}+\hat{{\mathcal{L}}}
_{N L}^{W}\label{pos1}
\end{eqnarray}
with
\begin{eqnarray}
\hat{{\mathcal{L}}} _{L}^{W}& =& \hat{\bar{\Psi}} ( i \gamma_{\mu} \partial^{\mu} -m_B ) 
\hat{\Psi} \nonumber\\ & + & \sum_{i=S,V,R,A...} s_i [ \frac{1}{2} (
\partial_{\nu}\hat{\phi}_i^{\mu} \partial^{\nu}\hat{\phi}_{i \mu}-m_i^2
\hat{\phi}_i^{\mu}\hat{\phi}_{i \mu})-g_i  \hat{\bar{ \Psi}} \Gamma_i^{\mu}\hat{\phi}_{i \mu}\hat{\Psi}],\label{pos2}
\end{eqnarray}
and
\begin{eqnarray}
\hat{{\mathcal{L}}} _{N L}^{W}=
 -\frac{1}{3} b_2 \hat{\varphi}^3-\frac{1}{4} b_3 \hat{\varphi}^4.\label{pos3}
\end{eqnarray}
where the meson contents can be seen in table~\ref{tab0}.

\begin{table}[h]
\centering
\begin{tabular}{|c||c|c|c|c|c|c|}
\hline Meson & $\hat{\phi}_{i \mu}$ (Field) & $g_i$ (Coupling 
Const) & $m_i$ (Mass) & $ \Gamma_i^{\mu}$ (Coupling terms)& $s_i$  \\\hline
$\sigma$ & $\hat{\varphi}(x)$ & $g_S$  & $m_S$  &1 & 1 \\\hline
$\omega$ & $\hat{V}_{\mu}(x)$ & $g_V$ &  $m_V$ & $\gamma_{\mu}$  & -1 \\\hline
$\rho$ & $\hat{\vec{R_{\mu}}}(x)$ & $\frac{g_R}{2}$ &  $m_R$ & $\gamma_{\mu}\vec{\tau}$  & -1 \\\hline
photon & $\hat{A}_{\mu}(x)$ & e &  0 & $\frac{1}{2}\gamma_{\mu}(1+\tau_3)$  & -1 \\\hline
\end{tabular}\\
\caption {The meson contents of equation (~\ref{pos2}).\label{tab0}}
\end{table}

For the $\phi_{i \mu}$, the fields used are $\hat{\varphi}(x)$, $\hat{V}_{\mu}(x)$, $\hat{\vec{R_{\mu}}}(x)$ and
$\hat{A}_{\mu}(x)$, denoting a scalar-isoscalar, a vector-isoscalar,
a vector-isovector and the electromagnetic field operators,
respectively. $g_i$ and $m_i$
are the coupling constants and masses of each field operator
$\hat{\phi}_{i \mu}$. We find that a difficulty
appears to determine the scalar field operator due to the
nonlinearities making it different from a simple Yukawa form. Some approximations have been done to overcome this
problem~\cite{ber,sugi,greco} but as far as we know there is
not yet an exact Hartree-Fock calculation including mesonic nonlinear terms.
Replacing  
\begin{eqnarray}
-\frac{1}{3} b_2 \hat{\phi}_s^3-\frac{1}{4} b_3 \hat{\phi}_s^4
&\longrightarrow& -\frac{1}{3}\beta_S  {(\hat{ \bar{
\Psi}}\hat{\Psi})}^3-\frac{1}{4} \gamma_S {(\hat{ \bar{
\Psi}}\hat{\Psi})}^4 \nonumber
\end{eqnarray}
we can  avoid the above problem and the standard definition for the linear field
operators can be used as~\cite{Lee}  
\begin{eqnarray}
\hat{\phi}_{i \mu}(x)&=&\hat{\phi}_{i \mu}^{0}(x)+  g_i \int d^4 y
D(x-y, m_i) \hat{\bar{ \Psi}} \Gamma_{ i\mu}\hat{\Psi},\nonumber\\
\hat{\phi}_{j \mu}^{0}(x)&=&\sum_{\alpha} (f_{\alpha j
\mu}(x)\hat{a}_{\alpha j}+ f_{\alpha j \mu}^{\dag}(x)
\hat{a}^{\dag}_{\alpha j}),\nonumber\\
\hat{\Psi}(x)&=& \sum_{\alpha}(\Psi_{\alpha}(x)\hat{b}_{\alpha }+\tilde{\Psi}_{\alpha}(x)\hat{d}^{\dag}_{\alpha }),\label{pos4}
\end{eqnarray}
where $\Psi_{\alpha}(x)$, $\tilde{\Psi}_{\alpha}$ and $f_{\alpha i
\mu}$ denote a nucleon, an antinucleon, and a meson wavefunction with $\alpha$
enumerating the states. The operators $\hat{a}_{\alpha j}$, $\hat{b}_{\alpha }$ and
$\hat{d}_{\alpha}$ annihilate a free meson, a nucleon and an antinucleon with
a momentum $k_{\alpha}$, with the conjugate operators the
corresponding creation operators. D(x,y) is the meson propagator
which is defined as a solution of
\begin{eqnarray}
(\partial_{\mu}\partial^{\mu} + m_i^2) D (x-y, m_i) =\delta ^{4}(x-y).\label{posi5}
\end{eqnarray}
The necessity
to introduce an extra density dependence in 
Dirac-Brueckner-Hartree-Fock calculations that allows a
simultaneous fit of the NN phase shift and the nuclear matter equilibrium
point~\cite{mach,brock} provides a physical motivation for 
this replacement.

Further on, we can study the connection between the Walecka and
point-coupling models by replacing the meson propagators with their low-order expansion plus
additional parameters $k_i$ as
\begin{eqnarray}
 -\int \frac{d^4
p}{(2\pi)^4}\frac{e^{-ip.(x-y)}}{p^2-m_i^2+i\epsilon}&\longrightarrow&  \frac{1}{m_i^2} {\delta}^{4}(x-y)-\frac{k_i}{m_i^4}\partial_{\mu}\partial^{\mu}{\delta}^{4}(x-y),
\label{posi6}
\end{eqnarray}   
to arrive at the Lagrangian density operator of the point-coupling (RMF-PC)~\cite{niko}
model in section 1. 

\subsection{Gordon decomposition of the exact equation of motion}
 After a straightforward Dirac matrix
algebra calculation we have from the Gordon decomposition of the exact
equation of motion that  
$A_1 +A_2+A_3$ =0, with 

\begin{eqnarray}
 A_1 &=& -\frac{1}{2
i}[(\partial^{\sigma}\bar{\Psi}_{\alpha})\Psi_{\alpha}-\bar{\Psi}_{\alpha}(\partial^{\sigma}\Psi_{\alpha})]\nonumber\\ 
&+&
\frac{1}{2}U_{\mu}[(\partial^{\mu}\bar{\Psi}_{\alpha})\gamma^{\sigma}\Psi_{\alpha}-\bar{\Psi}_{\alpha}\gamma^{\sigma}(\partial^{\mu}\Psi_{\alpha})]\nonumber\\&+&
\frac{1}{2}W_1 \partial_{\mu}[(\partial^{\mu}\bar{\Psi}_{\alpha})\gamma^{\sigma}\Psi_{\alpha}-\bar{\Psi}_{\alpha}\gamma^{\sigma}(\partial^{\mu}\Psi_{\alpha})]\nonumber\\&+&
\frac{1}{2}U_{\mu}^{\sigma}[(\partial^{\mu}\bar{\Psi}_{\alpha})\Psi_{\alpha}-\bar{\Psi}_{\alpha}(\partial^{\mu}\Psi_{\alpha})]\nonumber\\&+&
\frac{1}{2}W_2^{\sigma}\partial_{\mu}[(\partial^{\mu}\bar{\Psi}_{\alpha})\Psi_{\alpha}-\bar{\Psi}_{\alpha}(\partial^{\mu}\Psi_{\alpha})],\label{pos17}
\end{eqnarray}
\begin{eqnarray}
 A_2 &=& \frac{1}{2}\partial_{\mu}(\bar{\Psi}_{\alpha}
\sigma^{\mu\sigma}\Psi_{\alpha})- \frac{1}{2} i U_{\alpha \mu}\partial^{\mu}(\bar{\Psi}_{\alpha}
\sigma^{\alpha\sigma}\Psi_{\alpha})\nonumber\\&-& \frac{1}{2} i W_{2 \alpha}\partial^{\mu}\partial_{\mu}(\bar{\Psi}_{\alpha}
\sigma^{\alpha\sigma}\Psi_{\alpha})\label{pos18},
\end{eqnarray}
\begin{eqnarray}
 A_3 &=& 
\tilde{m}^*\bar{\Psi}_{\alpha}\gamma^{\sigma}\Psi_{\alpha}
+\tilde{V}^{\sigma}\bar{\Psi}_{\alpha} \Psi_{\alpha}.\label{pos19}
\end{eqnarray}
Let us define $\tilde{m}^* \equiv \bar{m}^* + \Delta m^*$
and $\tilde{V}^{\alpha} \equiv \bar{V}^{\alpha} + \Delta V^{\alpha}$,
where $\Delta m^*$ and $\Delta V^{\alpha}$ are the parts of
$\tilde{m}^*$ and  $\tilde{V}^{\alpha}$ of order $v^2$. 
Note that up to this point the treatment of exchange is still exact.
Now, if we apply the order-of-magnitude estimation of section \ref{sec3} to study
the magnitude in every term in $A_1$
through $A_3$ and assume that the terms of order $\geq$
$v^2$ are small and can be neglected, it is obvious that only the first term in  $A_1$ and  $A_2$ will survive and
the  $\Delta m^*$ and $\Delta V^{\alpha}$ contributions vanish in $A_3$. 
Therefore in this limit we will obtain Eq. (\ref{poses17}) of section (\ref{sec3}).

\subsection{The coupling constants of the approximate Lagrangian}
\label{Anhang:par}
\begin{subequations}
\begin{eqnarray}
  \tilde{\alpha}_S 
  &=& 
  \alpha_S-(\frac{1}{8} \alpha_S+\frac{4}{8}
  \alpha_V+\frac{12}{8} \alpha_{tV})
  - 
  2 m_B^2(\frac{1}{8} \delta_S+\frac{4}{8}
  \delta_V+\frac{12}{8} \delta_{tV})],
\label{pos38}
\\
  \tilde{\alpha}_V 
  &=&  
  \alpha_V+(-\frac{1}{8} \alpha_S+\frac{2}{8}
  \alpha_V+\frac{6}{8} \alpha_{tV})
  +2 m_B^2(-\frac{1}{8} \delta_S+\frac{2}{8}
  \delta_V+\frac{6}{8} \delta_{tV})
\nonumber\\
  &&
  +2 m_B^2(\frac{1}{8} \delta_S+\frac{4}{8}\delta_V
  +\frac{12}{8} \delta_{tV})
\\
  \tilde{\alpha}_{tS} 
  &=&  
  (-\frac{1}{8} \alpha_S-\frac{4}{8}
  \alpha_V+\frac{4}{8} \alpha_{tV})
  + 2 m_B^2(-\frac{1}{8} \delta_S-\frac{4}{8}
   \delta_V+\frac{4}{8} \delta_{tV})]
\\
  \tilde{\alpha}_{tV} 
  &=&  
  \alpha_{tV}+(-\frac{1}{8} \alpha_S+\frac{2}{8}
  \alpha_V-\frac{2}{8} \alpha_{tV})
  + 2 m_B^2(-\frac{1}{8} \delta_S+\frac{2}{8}
  \delta_V-\frac{2}{8} \delta_{tV})
\nonumber\\
  && 
  -2 m_B^2(-\frac{1}{8} \delta_S-\frac{4}{8}
  \delta_V+\frac{4}{8} \delta_{tV})]
\\
 \tilde{\delta}_S 
 &=& 
 \delta_S+ (\frac{1}{16} \delta_S+\frac{4}{16}
 \delta_V+\frac{12}{16} \delta_{tV})
\\
 \tilde{\delta}_V 
 &=& 
 \delta_V- (-\frac{1}{16} \delta_S+\frac{2}{16}
 \delta_V+\frac{12}{32} \delta_{tV})
\\
 \tilde{\delta}_{tS} 
 &=&  
 (\frac{1}{16} \delta_S+\frac{4}{16}\delta_V-\frac{4}{16} \delta_{tV})
\\
  \tilde{\delta}_{tV} 
  &=& 
  \delta_{tV}-(-\frac{1}{16} \delta_S+\frac{2}{16}
  \delta_V-\frac{4}{32} \delta_{tV})
\\
 \tilde{\theta}_T 
 &=& 
 4 m_B (\frac{1}{16} \delta_S+\frac{4}{16}
  \delta_V+\frac{12}{16} \delta_{tV})
\\
  \tilde{\theta}_{tT} 
  &=& 
  4 m_B (\frac{1}{16} \delta_S+\frac{4}{16}
  \delta_V-\frac{4}{16} \delta_{tV})
\label{pos47}
\end{eqnarray}
\begin{eqnarray}
 c_S &=& (\frac{1}{16} \delta_S+\frac{4}{16}
\delta_V+\frac{12}{16} \delta_{tV}),\nonumber\\
 c_V &=& (-\frac{1}{16} \delta_S+\frac{2}{16}
\delta_V+\frac{12}{32} \delta_{tV}),\nonumber\\ 
c_{tS} &=& (-\frac{1}{16} \delta_S+\frac{4}{16}
\delta_V-\frac{4}{16} \delta_{tV}),\nonumber\\
 c_{tV} &=& (-\frac{1}{16} \delta_S-\frac{2}{16}
\delta_V+\frac{4}{32} \delta_{tV}),\nonumber\\
\end{eqnarray}
\end{subequations}
\subsection{A complete mapping from Hartree-Fock to Hartree}
It has been shown that the usual ansatz for the point-coupling RMF model, when considered
within the Hartree-Fock approximation, leads to equations that deviate
from the structure of a Dirac-equation. The source for these
differences are the exchange terms with derivatives, leading to
densities that do not occur in the
standard Dirac-Hartree model. 

It should be mentioned, however, that the reason for these deviations is the
form of the Lagrange density operator, with which we start off.
We could as well choose a Lagrangian meant for a  strict Hartree treatment,
which would
lead to the same structure as we obtained it in this investigation. 
A similar situation holds for the Skyrme-Hartree-Fock theory.
The mapping can be achieved by adding terms that lead to the new densities
 in the Hartree-Fock case.
The Lagrangian density would have to be complemented by the following terms:
\begin{equation}
\Delta \hat{\mathcal L}^{derivative}_{new} = \sum_{j\in{\rm H}}\Big[ \frac{1}{2} i ~ \alpha_j ~ \hat{\mathcal A}_{j \mu} ~\hat {\mathcal
B}^{j \mu}+  \frac{1}{2} ~ \beta_j ~ \hat{\mathcal B}_{j
\mu} ~ \hat{\mathcal B}^{j \mu}+  \frac{1}{2} ~ \theta_j ~ \hat{\mathcal J}_{j}~
{\mathcal C}^{j} \Big],
\label{eq:new}
\end{equation}
here $\alpha_j$,  $\beta_j$ and $\theta_j$ are free parameters. This would include, however, the introduction of additional parameters, whereas in the HF case, no new parameters arise. The ansatz complemented by Eq. (\ref{eq:new}) leads to the same type of equation in both the Hartree and the Hartree-Fock approximations.

Though the mapping of the Hartree-Fock (and the corresponding effective Hartree) theory to a related strict Hartree theory can be done in principle, it would be at the cost of introducing and motivating new terms and additional coupling constants. If we also consider higher order terms \cite{ma}, a complete mapping to a Hartree theory would introduce many new parameters.
Experience will tell which ansatz for the effective interaction will prove
to be most adequate for a Hartree-Fock theory for finite nuclei.
\end{appendix}
\section*{Acknowledgements}
The authors would like to thank M.~Bender and T.~Cornelius for stimulating
discussions. A.S. greatfully acknowledges financial support from the
DAAD. This work was supported
in part by the Bundesministerium f\"ur Bildung und Forschung (BMBF),
Project No.\ 06 ER 808, by Gesellschaft f\"ur Schwerionenforschung
(GSI).


\begin{thebibliography}{50}
\bibitem{pgr} P.-G. Reinhard, M. Rufa, J. A. Maruhn, W. Greiner and J. Friedrich,
\Journal{\ZPA}{323}{13}{1986}
\bibitem{rufa} M. Rufa, P.-G. Reinhard, J. A. Maruhn, W. Greiner and
M.R. Strayer,
\Journal{\PRC}{38}{390}{1988}
\bibitem{pg} P.-G. Reinhard,
\Journal{\RPP}{52}{439}{1989}
\bibitem{sharma} M. M. Sharma, G. A. Lalazissis and P. Ring,
\Journal{\PLB}{317}{9}{1993}
\bibitem{Gambhir} Y.K. Gambhir, P. Ring and A. Thimet, 
\Journal{\AP}{198}{132}{1980}
\bibitem{Boersma} H.F. Boersma, 
\Journal{\PRC}{48}{472}{1993}
\bibitem{Lalazissis} G. A. Lalazissis et al.,
\Journal{\NPA}{608}{202}{1996}
\bibitem{Klemens}K. Rutz,
\Journal{}{ Dissertation}
{Frankfurt am Main}{1999}
\bibitem{Rutz2} K. Rutz et al.,
\Journal{\PRC}{56}{238}{1997}
\bibitem{bur}T. B{\"u}rvenich et al.,
\Journal{\EPJA}{3}{139}{1998}
\bibitem{cornel}T. Cornelius,
\Journal{}{ Diploma Thesis}
{Frankfurt am Main}{2001}
\bibitem{Kudling}L. Kudling,
\Journal{}{ Diploma Thesis}
{Frankfurt am Main}{2001}
\bibitem{Naza} W. Nazarewicz et al.,
\Journal{\PRC}{53}{740}{1996}
\bibitem{Bender3} M. Bender et al.,
\Journal{\PRC}{58}{2126}{1998}
\bibitem{burven}T. B{\"u}rvenich,
\Journal{}{ Dissertation}
{Frankfurt am Main}{2001}
\bibitem{buervi} T. B\"{u}rvenich, D. G. Madland, J. A. Maruhn, and 
     P. -G. Reinhard, 
\Journal{Phys. Rev. C 65 (2002) 044308}{}{}{2001} 
\bibitem{niko} B. A. Nikolaus, T. Hoch and D. G. Madland,
\Journal{\PRC}{46}{1757}{1992}
\bibitem{sur} J. J. Rusnak and R. J. Furnstahl,
\Journal{\NPA}{627}{95}{1997}
\bibitem{nley} J. K. Zhang, Y. Jin and D. S. Onley,
\Journal{\PRC}{48}{2697}{1993}
\bibitem{schmd} R. N. Schmid, E. Engel and R. M. Dreizler,
\Journal{\PRC}{52}{164}{1995}; \Journal{\PRC}{52}{2804}{1995}; \Journal{\FP}{27}{1257}{1997}
\bibitem{Lee} S. J. Lee,
\Journal{ } {Dissertation }
{Yale University}{1986}
\bibitem{ber} P. Bernados et al.,
\Journal{\PRC}{48}{2665}{1993}
\bibitem{sugi} S. Sugimoto, K. Sumiyoshi and H. Toki,
\Journal{\PRC}{64}{054310}{2001}
\bibitem{greco} C. Greco  et al.,
\Journal{}{nucl-th}{0010092}{2000}
\bibitem{mach} R. Machleidt,
\Journal{Adv.~ Nucl.~ Phys}{19}{189}{1989}
\bibitem{brock} R. Brockmann and R. Machleidt,
\Journal{\PRC}{42}{1990}{1985}
\bibitem{ma} J. A. Maruhn, T. B{\"u}rvenich, D. G. Madland,
\Journal{Journal of Comput. Phys}{238}{169}{2001}
\bibitem{GM} W. Greiner, B. M{\"u}ller,
\Journal{, Gauge Theory of Weak Interaction, } {Springer-Verlag}
{Heidelberg}{1989}
\bibitem{greiner} W. Greiner, J. Reinhard,
\Journal{Quantum Electrodynamics, } {Springer-Verlag}
{Heidelberg}{1997}
\bibitem{ben} M. Bender,
\Journal{} {Dissertation }
{Frankfurt am Main}{1997}
\bibitem{mach2} R. Machleidt, K. Holinde, C. Elster,
\Journal{\PRpt}{C149}{1}{1987}
\bibitem{otna} A.~Sulaksono, T.~B{\"u}rvenich, J.~A.~Maruhn, W.~Greiner, and
P.-G.~Reinhard, in preparation
\end{thebibliography}
\end{document}